\documentclass[journal]{IEEEtran}

\usepackage{cite}

\ifCLASSINFOpdf
  \usepackage[pdftex]{graphicx}
  \graphicspath{{images/}}
  \DeclareGraphicsExtensions{.pdf,.jpg,.png}
\else
\fi
\usepackage{color}
\usepackage{amsmath}
\ifCLASSOPTIONcompsoc
 \usepackage[caption=false,font=normalsize,labelfont=sf,textfont=sf]{subfig}
\else
 \usepackage[caption=false,font=footnotesize]{subfig}
\fi

\usepackage{multirow}

\usepackage[ruled,linesnumbered]{algorithm2e}
\usepackage{soul}

\begin{document}
%
\title{End-to-End Asynchronous Traffic Scheduling in Converged  
5G and Time-Sensitive Networks}
%
%
%

\author{Jiacheng~Li, 
        Yongxiang~Zhao, 
        Chunxi~Li,
        Zonghui~Li,
        Kang~G.~Shin,~\IEEEmembership{Life~Fellow,~IEEE,}
        and Bo~Ai,~\IEEEmembership{Fellow,~IEEE}
\thanks{Manuscript received ; revised . This work was supported in part by the National Key Research and Development Project under Grant 2022YFB3303702, and in part by the National Natural Science Foundation of China under Grant 62002013 and Grant  62272034. (\textit{Corresponding author: Zonghui Li})}
\thanks{Jiacheng Li, Yongxiang Zhao, Chunxi Li and Bo Ai are with the School of Electronic and Information
Engineering, Beijing Jiaotong University, Beijing 100044,
China (e-mail: 21120075@bjtu.edu.cn; yxzhao@bjtu.edu.cn; chxli1@bjtu.edu.cn; boai@bjtu.edu.cn).}
\thanks{Zonghui Li is with the School of Computer and Information
Technology, Beijing Jiaotong University, Beijing 100044, China (e-mail: lizonghui@bjtu.edu.cn).}
\thanks{Kang G. Shin is with the Department of 
Electrical Engineering and Computer Science, University of Michigan, Ann Arbor, MI 48109-2121, USA (e-mail: kgshin@umich.edu).}
}

%
%

\markboth{Journal of \LaTeX\ Class Files,~Vol.~14, No.~8, August~2015}%
{Shell \MakeLowercase{\textit{et al.}}: Bare Demo of IEEEtran.cls for IEEE Journals}
%



\maketitle

\begin{abstract}
As required by Industry 4.0, companies will move towards flexible 
and individual manufacturing. To succeed in this transition, 
convergence of 5G and time-sensitive networks (TSN) is the 
most promising technology and has thus attracted considerable
interest from industry and standardization groups. 
However, the delay and jitter of end-to-end (e2e) transmission 
will get exacerbated if the transmission opportunities are missed 
in TSN due to the 5G transmission jitter and the clock skew 
between the two network systems. To mitigate this phenomenon, 
we propose a novel asynchronous access mechanism (AAM) that isolates the 
jitter only in the 5G system and ensures zero transmission jitter 
in TSN. We then exploit AAM to develop an e2e asynchronous traffic scheduling model for coordinated allocation of resources for 5G 
and TSN to provide e2e transmission delay guarantees for 
time-critical flows. 
The results of our extensive simulation of AAM on OMNET++ corroborate 
the superior performance of AAM and the scheduling model.
\end{abstract}


%
\IEEEpeerreviewmaketitle

\section{Introduction}
\label{introduction}
%
%
%
%
\IEEEPARstart{M}{ore} and more industrial use-cases 
require low-delay communication to enable production automation 
\cite{GarcaMorales2019LatencySensitive5R, Wollschlaeger2017TheFO}, 
such as the control of cooperating robots 
\cite{Arnim2020TSNbasedCI} and Automated Guided Vehicles (AGVs) 
\cite{9453651}. Real-time networks serve as the foundation for 
providing deterministic and reliable transmission for such 
industrial time-sensitive communication. According to the 
prevalent transmission media, industrial real-time networks can
be characterized as a combination of wired and wireless networks. 

Current industrial real-time networks are still dominated by 
wired technologies \cite{Industrial_Ethernet} such as CAN, 
PROFIBUS, EtherCAT, and TTEthernet. However, these technologies 
either have low data-rate \cite{Raagaard2018OptimizationAF},
or are manufacturer-specific, thereby making them incompatible 
with each other \cite{Deng2022ASO}. Consequently, IEEE 802.1 
Time-Sensitive Networking (TSN) \cite{TSNGROUP}, which 
defines a variety of mechanisms such as time synchronization and 
traffic shaping to guarantee deterministic transmission, has been 
a promising solution for future unified industrial 
networks in the wired environment \cite{LoBello2019APO}.

However, such a wired technology is only suitable for production 
facilities that are static and have long life cycles
\cite{Khoshnevisan20195GIN},  thus lacking the flexibility and 
modularity \cite{Ginthr2019AnalysisOM, Atiq2022WhenI8} required by 
Industry 4.0. The industry and academia, therefore, turn to 
wireless technology to achieve this goal
\cite{Kehl2022PrototypeO5,Khoshnevisan20195GIN}.

The fifth-generation cellular network technology (5G), which considered 
industrial use-cases from the very beginning, has been considered as a 
key technology for two reasons. First, its ultra-reliable 
low-delay communication (URLLC) feature provides a flexible frame 
structure,  faster signaling, etc., to support low-delay 
communication \cite{HamidiSepehr20215GUE}. Second,
3GPP has completed the integration architecture of 5G system (5GS) and 
TSN \cite{3gpp.23.501}, which includes additional functionalities, 
such as the Network-side TSN translator (NW-TT) for user plane 
function (UPF) to access TSN.

However, simple integration of 5GS and TSN may significantly exacerbate 
the delay and jitter in packet delivery. First, assuming TSN 
uses the time-aware shaper (TAS), each packet from 5GS must arrive 
at the gateway before the start of its "prepared" 
transmission opportunity in TSN, which we call the {\em time-triggered 
access mechanism} (TAM). Second, it is difficult for 5GS to 
successfully deliver a packet before the start of its transmission 
opportunity owing to (1) the transmission jitter caused by wireless 
transmission in 5GS and (2) the clock skew between 5GS and TSN. 
Thus, if a packet output by 5GS misses its transmission opportunity 
in TSN, it has to wait for the opportunity assigned to the next packet. 
Since the interval between TSN transmission opportunities is
much larger than (1) and (2), 
missing a transmission opportunity will enlarge the e2e 
delay and jitter significantly.

To mitigate this delay and jitter enlargement, we propose a novel {\em asynchronous access mechanism} (AAM) to bridge 5GS and TSN.
By decoupling the transmission interaction between 5GS and TSN, the 
AAM ensures zero transmission jitter in TSN at the cost of the wired 
transmission bandwidth, no matter how randomly the packets arrive 
at TSN from 5GS. 
Using the AAM, we then build an e2e {\em asynchronous traffic scheduling 
model} (ATSM) to allocate the resources of TSN and 5GS. 
We then validate the performance of AAM and ATSM via simulation 
on OMNET++\cite{OMNET}.

The remainder of this paper is organized as follows. 
Section \ref{relatedWork} discusses the related work while
Section \ref{asynchronous access mechanism} introduces the delay 
and jitter enlargement in the context of TAM and mitigates it
with the AAM. Section \ref{asynchronous traffic scheduling model} 
proposes the ATSM based on the AAM. Section \ref{PerformanceEvaluation} 
evaluates the performance of the AAM and the ATSM. 
Finally, this paper concludes with Section \ref{conclusion}.

\section{Related Work}
\label{relatedWork}

The traffic scheduling for the converged network of 5G and TSN is still 
in its infancy and can be treated as either separate or joint 
scheduling depending on whether the resources in 
the two networks are allocated cooperatively or not. 

In the case of separate scheduling, the e2e delay requirement 
of a time-critical flow is first divided into 5GS and TSN delay
budgets, and then the two networks schedule their own network 
resources independently according to their own delay budget. 
\cite{Larraaga20235GCG} scheduled multiple resource blocks (RBs) 
for multiple packets of time-critical flows to solve 
the periodic resource conflict in 5GS. 
\cite{Cai2023DynamicQM} dynamically mapped the QoS requirements 
obtained from time-sensitive communication assistance 
information (TSCAI) to 5G QoS Identifier (5QI), and added dynamic 
RBs to pre-allocated RBs if the latter was insufficient for
data transmission. \cite{Yang2022TrafficSF} maximized the ratio of 
eMBB traffic throughput to URLLC traffic delay given the flow 
information from TSCAI. Most of the related work was devoted to studying 
how to achieve the resource allocation in 5GS given the 5GS delay 
budget without considering how to allocate delay budgets to TSN and 5GS.

Joint scheduling manages the use of resources cooperatively in TSN 
and 5GS from an e2e perspective. 
The authors of \cite{Wang2023ReinforcementLP} proposed a unified 
scheduling model to allocate appropriate time slots 
for time-critical flows using the cyclic queuing and forwarding (CQF) 
mechanism and proposed a reinforcement learning-based algorithm to search 
for the optimal scheduling solution. 
\cite{2023DecAgeDF} maintained the freshness of information by 
scheduling sampling and transmission decisions cooperatively to 
reduce energy consumption. The flow pattern output from TSN to 
5GS was found in \cite{Ginthr2019AnalysisOM} to have a significant 
impact on the capacity and delay in 5GS, and \cite{Ginthr2020EndtoendOJ} 
proposed an e2e-optimized scheduling framework to jointly schedule the 
resources of the two network systems. 
All these works assumed that 5GS and TSN were synchronized. 

However, we find the transmission interaction between 5GS and TSN 
possibly resulting in the delay and jitter enlargement as 
mentioned in Section \ref{introduction}. To mitigate this, we propose 
the AAM, which enables the converged network to run in asynchronous 
conditions. Furthermore, we propose the ATSM based on the AAM to 
coordinate the allocation of 5GS and TSN resources.

\section{asynchronous access mechanism}
\label{asynchronous access mechanism}

We first introduce the application scenario and the e2e delay and 
jitter enlargement in the context of TAM. The basic ideas 
and implementation details of AAM are then presented to 
mitigate this problem. 

\subsection{Application Scenario}
\label{application scenario}

\begin{figure}[!t]
\centering
\includegraphics[width=3.5in]{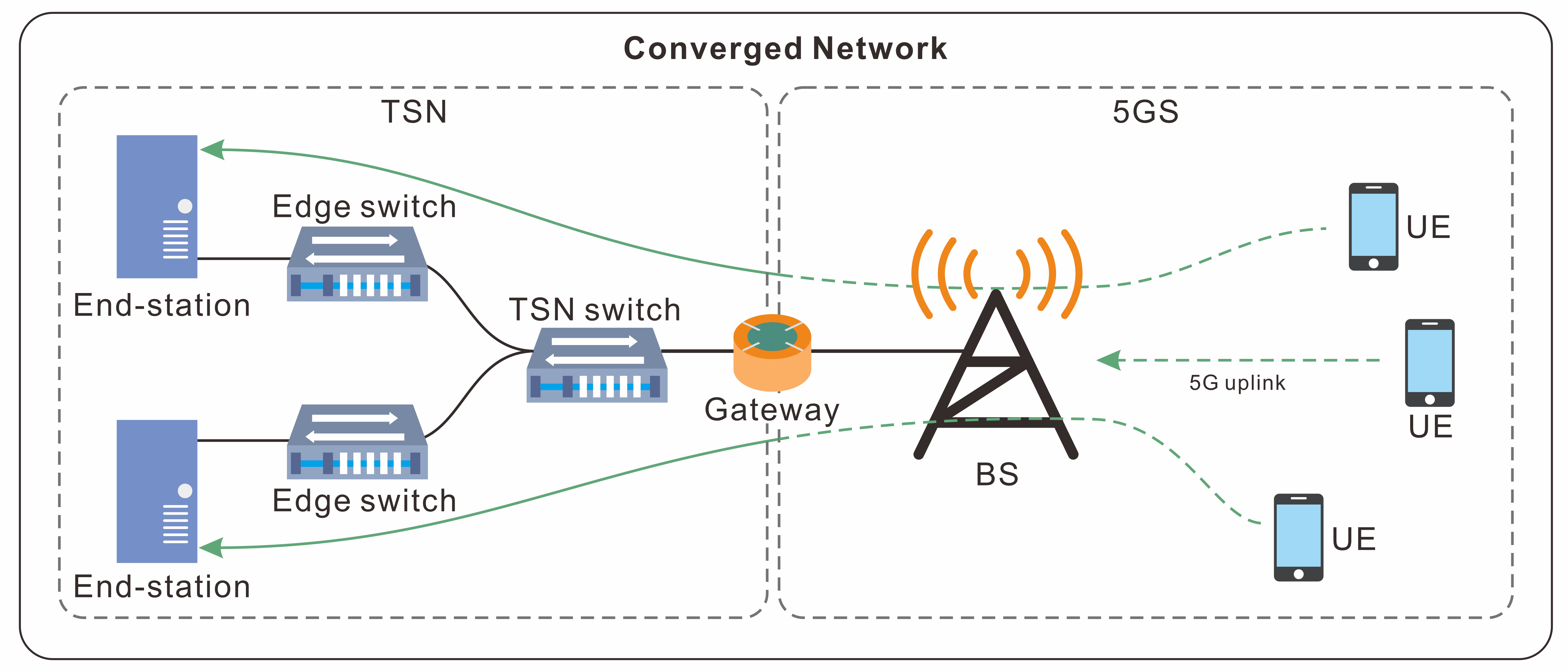}
\caption{The converged network that integrates 5GS and TSN.}
\label{fig:NetworkModel}
\end{figure}

Fig.~\ref{fig:NetworkModel} shows an application scenario
in which a TSN and a 5GS are connected via a gateway (GW). 
TSN is composed of end-stations (ESs) and TSN switches 
while 5GS is composed of user equipment (UEs) and a base 
station (BS). All the flows under consideration originate from UEs 
in 5GS to the ESs in TSN and belong to time-triggered (TT) flows. 
Specifically, the source of a TT flow periodically generates 
fixed-length packets. Each TT flow is subject to
the e2e delay requirement for packet delivery. 
 
Additionally, we assume that the clocks in TSN and 5GS are 
synchronous within their respective domains but can be 
asynchronous between different domains. Each TSN switch 
adopts TAS on each port.
We also assume that 5GS uses the FDD duplex scheme where 
downlink and uplink transmissions use different frequencies 
and all UEs share a common 5G uplink.

\subsection{Delay and Jitter Enlargement}
\label{sec:delayJitterEnlargement}
\begin{figure*}[!t]
    \centering

    \subfloat[Issue caused by clock skew.]{
            \includegraphics[width=3in]{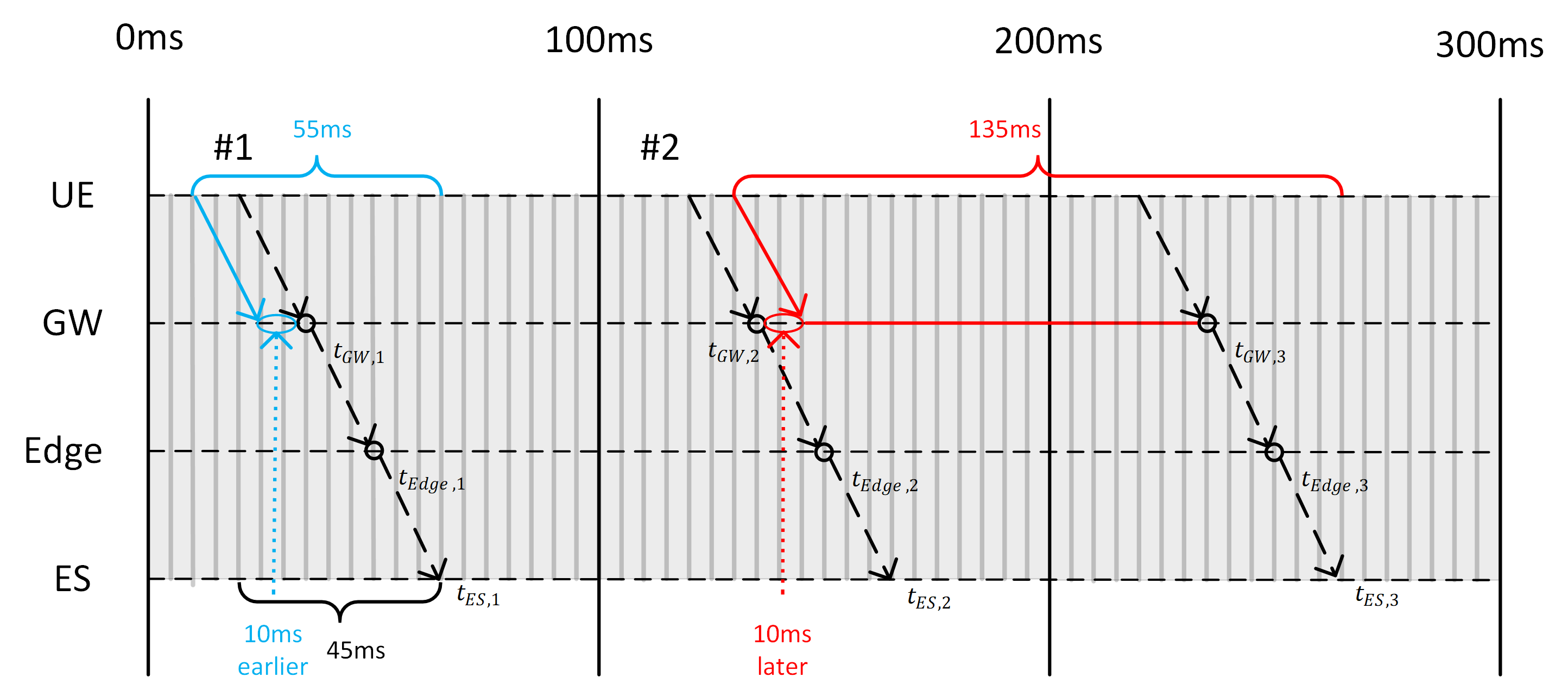}
            \label{fig:AAM-1}
    }
    \hfill
    \subfloat[Issue caused by the 5G transmission jitter.]{
            \includegraphics[width=3in]{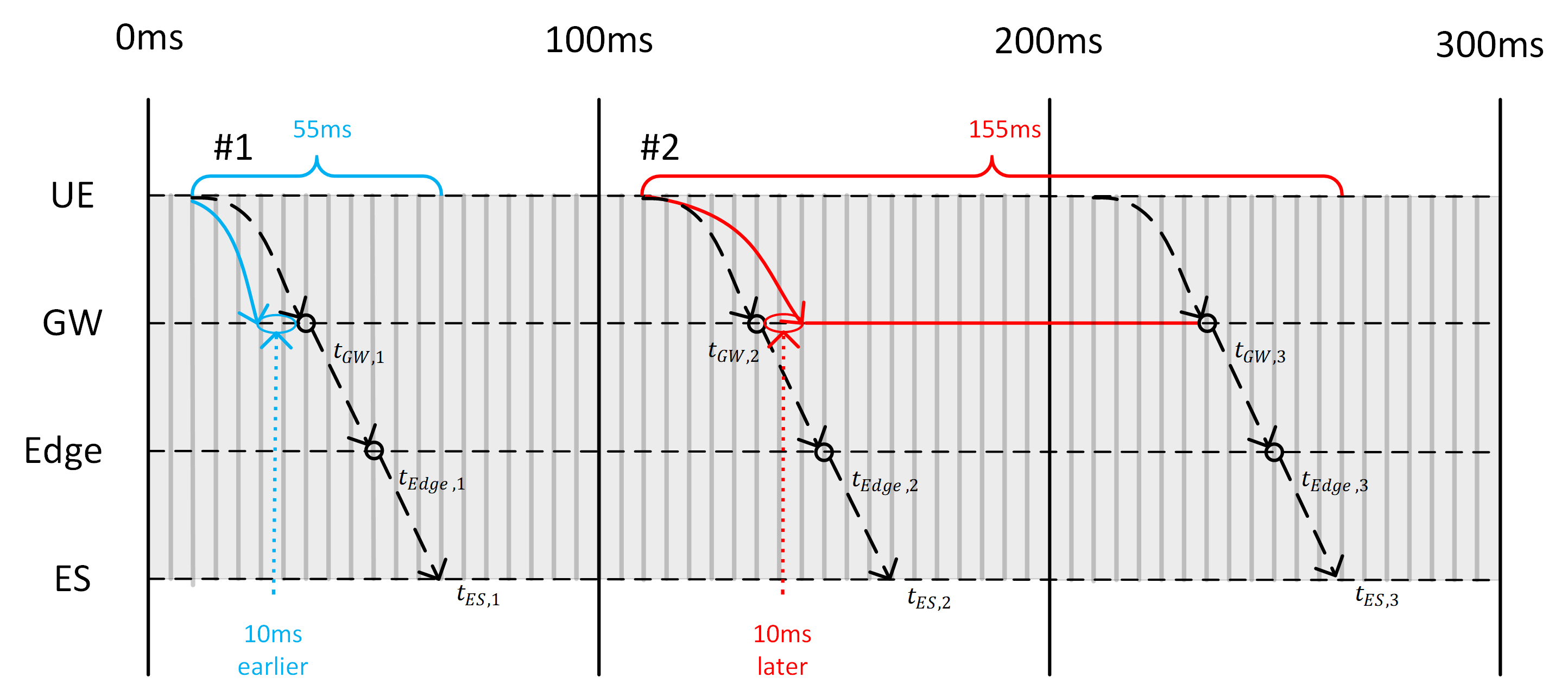}
            \label{fig:AAM-2}
        
    }

    \subfloat[AAM solution of the issue in (a).]{
            \includegraphics[width=3in]{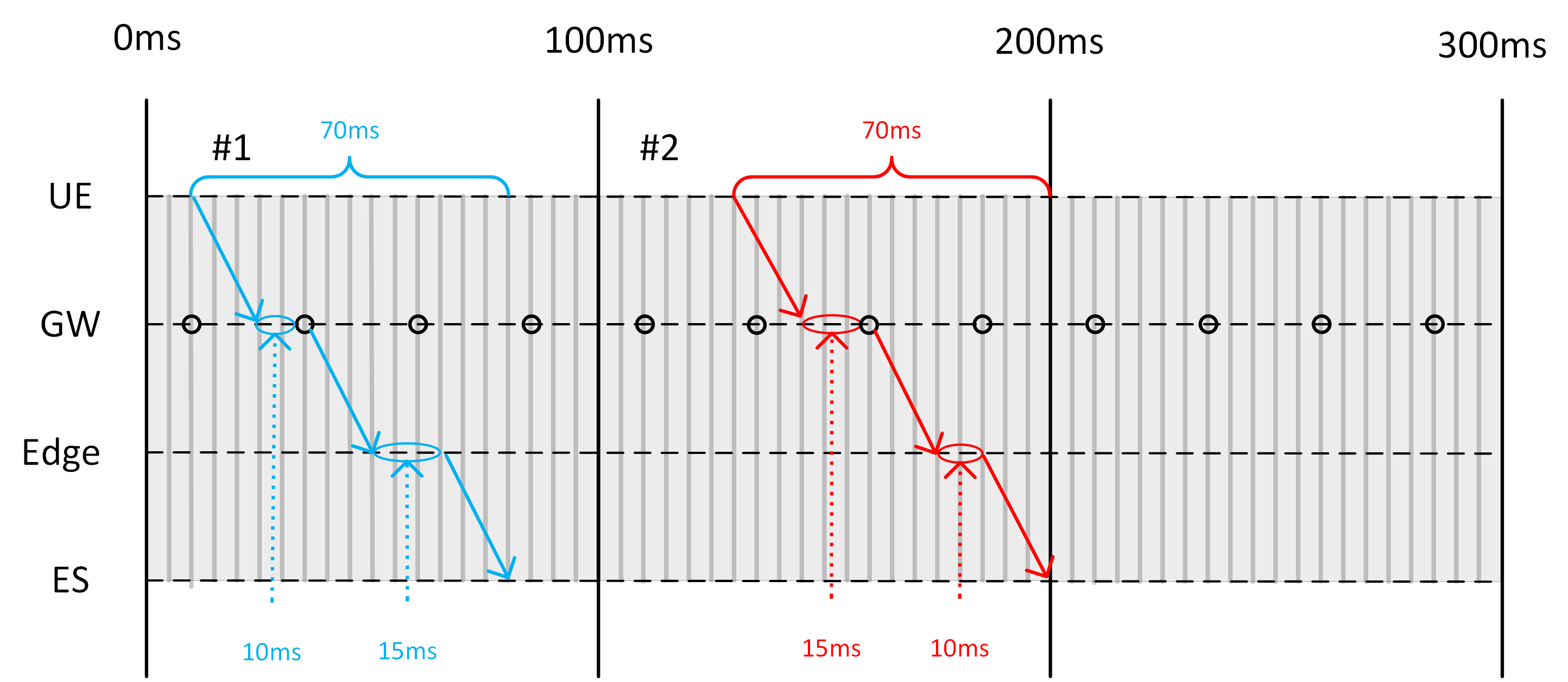}
            \label{fig:AAM-3}
       
    }
    \hfill
    \subfloat[AAM solution of the issue in (b).]{
            \includegraphics[width=3in]{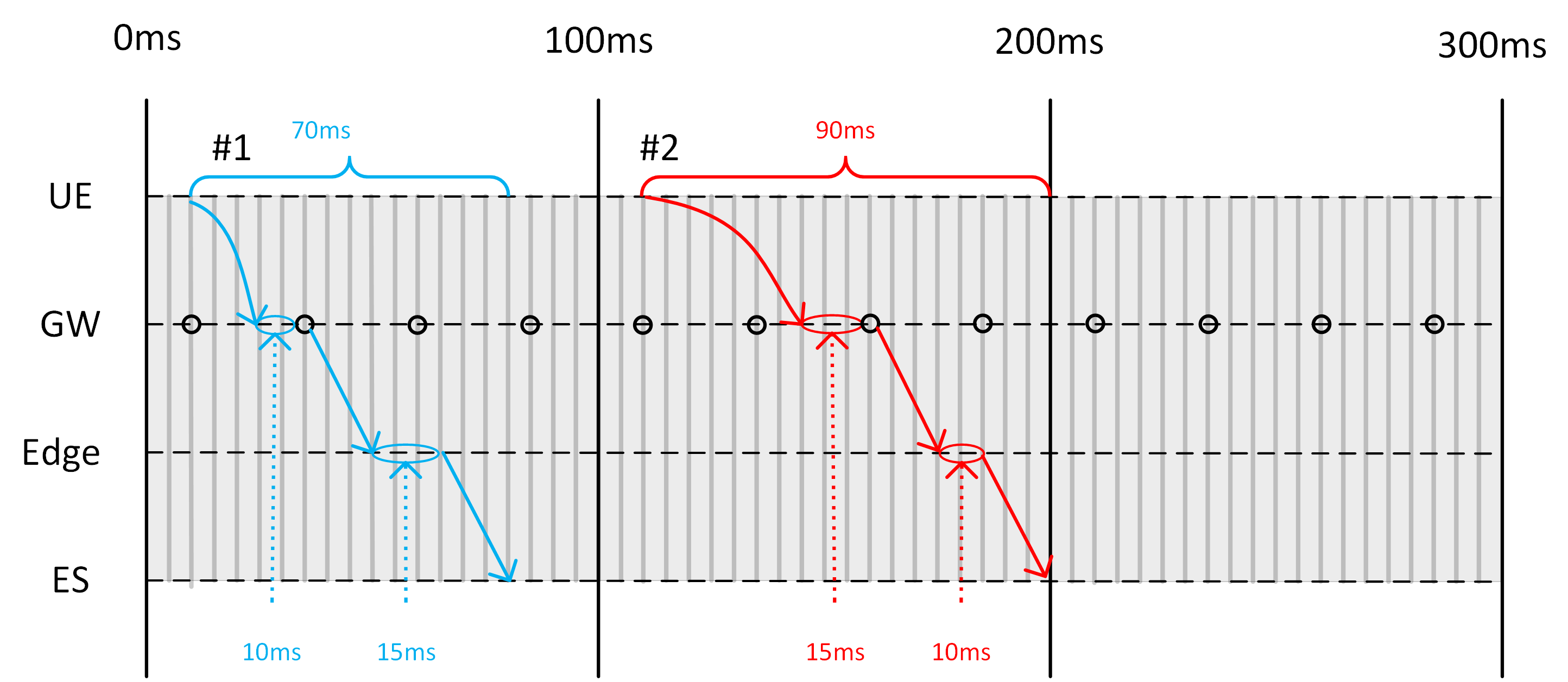}%
            \label{fig:AAM-4}
        
    }

\caption{Issues caused by the 5G transmission jitter and 
the clock skew in the context of TAM and their mitigation with AAM.}  
\label{fig:detailedTwoMechanism}
\end{figure*}

Using an example, we show below that the simple integration of 5GS 
and TSN may enlarge e2e delay and jitter, where the 
simple integration means TAS's reservation of a transmission 
opportunity for each frame of a TT flow on each link on its route.

In Figs.~\ref{fig:AAM-1}--\ref{fig:AAM-4}, the UE in 5GS sends 
a packet every $100ms$ to the ES in TSN via the GW and edge 
switch (Edge).  We will use Figs.~\ref{fig:AAM-1} and 
\ref{fig:AAM-2} to show the enlargement of delay and jitter 
caused by the clock skew between 5GS and TSN and 
the 5G transmission jitter, respectively.

As shown in Fig.~\ref{fig:AAM-1}, in an ideal situation --- i.e., 
there is no 5G transmission jitter or clock skew between 5GS and 
TSN --- GW and Edge reserve a transmission opportunity every 
$100ms$ (= the flow period) on their output port, which is 
shown as black hollow circles, and the black dashed arrows 
depict the ideal transmission processes. Each packet experiences 
a constant delay of $45ms$ and the e2e jitter of 0. 
However, considering the clock skew between the two systems, 
the packets may arrive at TSN earlier (depicted by the blue line) 
or later (depicted by the red line) from the perspective of TSN. 
We assume the maximum clock skew is $10ms$, i.e., the clock of 
TSN may be at most $10ms$ earlier or later than that of 5GS. 
On the one hand, packet \#1 arrives $10ms$ earlier than the
prepared transmission opportunity and has to wait for it, 
leading to the actual e2e delay of $55ms$. On the other hand, 
packet \#2 arrives $10ms$ later than the prepared transmission 
opportunity and must wait for the opportunity in the next 
period, leading to the actual e2e delay of $135ms$. 
Actually, the experienced e2e delay can be as low as $45ms$ and 
as large as $145ms$, creating a deviation of $100ms$ which 
equals the flow period. Thus, a small skew between the clocks 
of the two systems may lead to a significantly larger e2e 
delay and jitter than scheduled.

Similarly, as shown in Fig.~\ref{fig:AAM-2}, we assume that
the clocks are synchronized, the delay budget in 5GS is 
estimated to be the average delay over a period of time, 
and the transmission opportunities prepared by TSN are scheduled 
by reserving the delay budget of 5GS. The transmission jitter of 
5GS is $10ms$, meaning packets may arrive at most $10ms$ quicker 
or slower than the ideal arrival time calculated by the delay 
budget. Packet \#1 experiences an e2e delay of $55ms$ while 
packet \#2 experiences $155ms$ and their deviation shown in 
the figure is thus $100ms$, which also equals the flow period. 
A small 5G transmission jitter may also lead to a significantly 
larger e2e delay and jitter than scheduled.

In fact, the e2e delay and jitter enlargement is caused by 
missing TSN transmission opportunities. To mitigate this 
enlargement, we propose the AAM to decouple the interaction 
between 5GS and TSN transmissions.

\subsection{Basic Ideas of the AAM}
\label{sec:basicIdeasofAAM}

The basic ideas of the AAM are to (1) hold the packets at the 
edge switches to shape the resident time in TSN to be constant 
and (2) over-provision the transmission opportunities to 
reduce delay. 

First, a holding process is introduced in the edge switch. Let $T_i$
be the period of the transmission opportunities (not necessarily equals 
the flow period as will be discussed later). As shown in 
Fig.~\ref{fig:AAM-3}, $T_i=25ms$. Packets \#1 and \#2 experience similar 
transmission processes: (1) waiting at the GW, (2) transmitted from GW to 
Edge, (3) held at the Edge, and (4) transmitted from Edge to ES, and the 
corresponding delays are denoted as $f_i.wait$, $D^{TSN}_{i,(1)}$, 
$f_i.hold$ and $D^{TSN}_{i,(2)}$. The delay in TSN can be represented as:
\begin{gather}
    D^{TSN}_i = f_i.wait + D^{TSN}_{i,(1)} + f_i.hold + D^{TSN}_{i,(2)},
    \label{eq:e2edelayTSNOrigin}
\end{gather}
where $D^{TSN}_{i,(1)}$ and $D^{TSN}_{i,(2)}$ are both constant assuming 
TSN is well organized. It is easy to get $f_i.wait\in [0, T_i)$ 
so the holding time is for compensating the waiting time to be $T_i$, i.e., 
$f_i.wait+f_i.hold = T_i$. As can be seen from Fig.~\ref{fig:AAM-3}, 
packet \#1 waits for $10ms$ and is held for $15ms$, and 
packet \#2 waits for $15ms$ and is held only for $10ms$. 
This way, the packets' resident time in TSN 
is fixed and can be represented as:
\begin{align}
    D^{TSN}_i = T_i + D^{TSN}_{i,(1)} + D^{TSN}_{i,(2)}.
    \label{eq:delayTSN}
\end{align} 
So, all the packets of the flow experience the same delay in TSN no 
matter when the packets are delivered by 5GS, and the transmission 
coupling between the two network systems is eliminated, hence mitigating 
the delay and jitter enlargement.  As shown in Fig.~\ref{fig:AAM-3}, both 
packets experience the same e2e delay of $70ms$ regardless of the clock 
skew between the two network systems. 

However, in many industrial use-cases \cite{Raagaard2018OptimizationAF} 
the delay requirement is no larger than the period. If $T_i=f_i.period$, 
the delay requirement cannot be satisfied according to 
Eq.~\eqref{eq:delayTSN}. 
This way the transmission opportunities are over-provisioned to 
reduce delay, i.e., $T_i$ should be less than $f_i.period$. 

Similarly, Fig.~\ref{fig:AAM-4} illustrates how the AAM can help 
mitigate the issues presented in Fig.~\ref{fig:AAM-2}. 
Moreover, the e2e transmission jitter is isolated only in 5GS.

Note that the parameter $T_i$ should be carefully determined to  
balance between the resource usage and the delay in TSN, and between 
the delay budgets in TSN and 5GS for the following reasons:
\begin{itemize}
    \item $T_i$ allows for a trade-off between usage of TSN resource and 
    the delay in TSN. According to Eq.~\eqref{eq:delayTSN}, the smaller 
    $T_i$, the lower the delay in TSN. However, a smaller $T_i$ means 
    more TSN resource usage for a single flow as shown in 
    Figs.~\ref{fig:AAM-1} and \ref{fig:AAM-3}.
    \item $T_i$ balances the delay budgets in 5GS and TSN. 
    The e2e delay is composed of 5G and TSN parts:
    \begin{align}
    D_i^{e2e}&=D_i^{5GS}+D_i^{TSN}\nonumber\\
    &=D_i^{5GS}+T_i + D^{TSN}_{i,(1)} + D^{TSN}_{i,(2)}. 
    \label{eq:e2eDelay}
    \end{align}
    Given the e2e delay requirement of a flow, a larger $T_i$ will 
    decrease the delay budget allocated to 5GS which will affect 
    the performance of 5GS.
\end{itemize}

It is, therefore, necessary to establish a comprehensive model to 
jointly schedule the resources of 5GS and TSN when the AAM is 
applied; this will be discussed in 
Section \ref{asynchronous traffic scheduling model}.

\subsection{Implementation of the AAM}
The AAM is composed of the following two parts.
\begin{itemize}
    \item The gateway caches the received packet of $f_i$ until 
    the next transmission opportunity, attaches the waiting time 
    $f_i.wait$ and sends it to the destination.
    \item The edge switch of the last hop holds the packet of $f_i$ 
    for time duration $f_i.hold$ and then delivers it to the final 
    destination, where $f_i.hold+f_i.wait=T_i$.
\end{itemize}
The implementation details in the gateway and the edge switch are described in Algorithms \ref{algorithm:gateway} 
and \ref{algorithm:edge}, respectively.

\begin{algorithm}
\SetAlgoNoLine
function onReceivingPacketFrom5GS($f_i.packet$):

\Indp
$f_i.buffer=f_i.packet.getPayload()$;

$f_i.receivedTime=now()$;

\Indm function onTransmissionOpportunity($f_i.timer$):

\Indp
$resetTimer(f_i.timer)$;

\eIf{$f_i.buffer$ is empty}
    { return\;}
    {
     new $f_i.packet$\;
    
    $f_i.packet.setPayload(f_i.buffer)$\;
    
    $f_i.packet.setWait( now()-f_i.receivedTime)$\;
    
     send($f_i.packet$)\;
     $f_i.buffer.reset()$\;
    }
\Indm

\caption{AAM at the gateway}

\label{algorithm:gateway}
\end{algorithm}

\begin{algorithm}
function onReceivingPacket($f_i.packet$):

\Indp delay($f_i.T-f_i.packet.getWait()$)\;

 send($f_i.packet$)\;

\caption{AAM at the edge switch}
\label{algorithm:edge}
\end{algorithm}

In Algorithm \ref{algorithm:gateway}, there are two functions in 
the gateway. In the first function onReceivingPacketFrom5GS($f_i.packet$), once the 
packet of flow $f_i$ is received, the gateway caches the packet in the 
buffer $f_i.buffer$ prepared for this flow (Line 2). If there is already 
a packet in the buffer, the old packet will be replaced by the new 
packet. Then, the reception time of the packet is recorded (Line 3). 
In the second function onTransmissionOpportunity ($f_i.timer_i$), the 
function takes a timer as an input parameter. Since a timer is set for 
each flow $f_i$ to inform its transmission, the function first 
resets the timer for the next transmission opportunity for this flow 
(Line 5). It then checks whether the buffer of this flow is empty. 
On the one hand, the function does nothing if the corresponding buffer 
is empty. On the other hand, it attaches the waiting time to this 
packet and sends it before clearing the buffer if the buffer is 
not empty (Lines 6-13).

There is only one function in Algorithm \ref{algorithm:edge}. 
When the edge switch receives the packet $f_i.packet$, it delays 
the packet for the amount of time $f_i.T-f_i.packet.getWait()$ 
which is consistent with the process mentioned before (Line 2). 
The packet is then delivered to the destination after the 
delay (Line 3).

\begin{figure}[!t]
\centering
\includegraphics[width=3.5in]{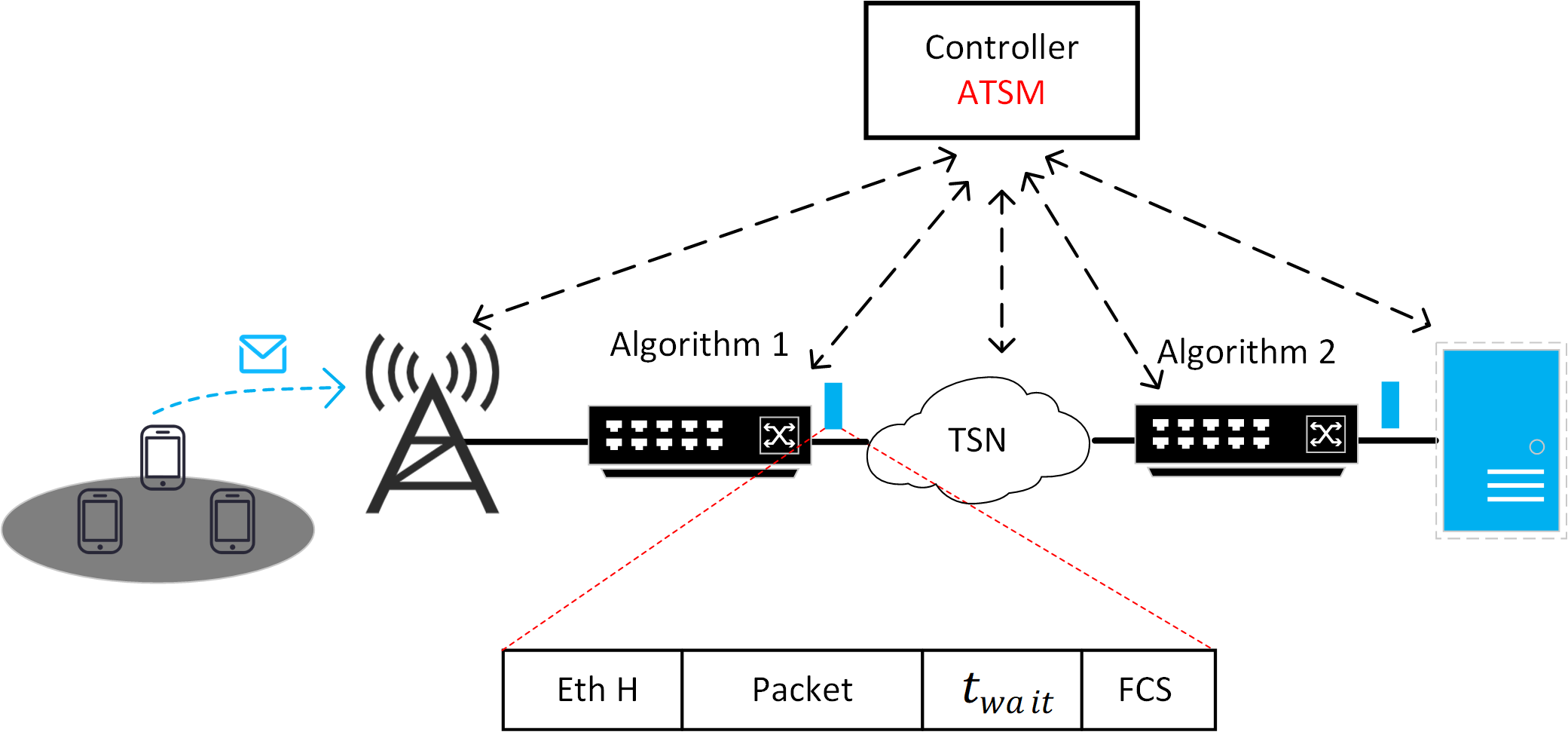}
\caption{An example of the deployment of the AAM.}
\label{fig:AAMDeployment}
\end{figure}

As shown in Fig.~\ref{fig:AAMDeployment}, AAM can be easily 
integrated into the 5GS-TSN architecture proposed in 
\cite{3gpp.23.501}. The base station complies with the 5G 
Network standard. The two black boxes implement 
Algorithm~\ref{algorithm:gateway} and 
\ref{algorithm:edge}, respectively, and they are connected by a 
traditional TSN network. The frame’s waiting time at the left 
black box is incorporated into the frame payload and the right black
box can extract the time from the received frames. The ATSM is 
implemented in the controller, whose detailed structure can be 
found from \cite{3gpp.23.501}. The controller collects the 
state of the converged network and flows' requirements, and then 
calculates the settings of 5GS, TSN, and the two black boxes. 

\section{asynchronous traffic scheduling model}
\label{asynchronous traffic scheduling model}
We now formally present ATSM to schedule the resources in the 
converged network based on AAM. The network and traffic models are 
introduced first and the scheduling model is presented to 
formulate the scheduling problem. We then derive the formal network 
constraints and the objective function. Finally, a linearization 
method is introduced to transform the scheduling problem 
to an Integer Linear Programming problem.

\subsection{Network and Traffic Models}
The wired TSN network is defined as an undirected graph $G(V, E)$, 
where $V$ is the set of nodes such as ESs, the gateway, and TSN 
switches, and $E$ is the set of physical full-duplex links in TSN. 
Specifically, each physical link contains two dataflow 
links \cite{Steiner2010AnEO}. Let $L^{TSN}$ denote the set of 
dataflow links in TSN, then
\begin{equation}
\forall v_1,v_2\in V ,(v_1,v_2 )\in E :[v_1,v_2] ,[v_2,v_1] \in L^{TSN},
\end{equation}
where "$(v_1,v_2)$" denotes the physical link between nodes $v_1$ and 
$v_2$, "$[v_1,v_2]$" and "[$v_2,v_1]$" are the corresponding 
directed dataflow links.

The set of dataflow links in 5GS is denoted as $L^{5GS}$. 
The entire set of dataflow links in the converged network can 
then be represented as $L = \{L^{TSN}, L^{5GS}\}$.

Let $F$ be the set flows under consideration. Each flow $f_i \in F$ 
is identified as:
\begin{equation}
\forall f_i\in F: <f_i.period,f_i.length,f_i.delay>,
\end{equation}
where $f_i.period$ denotes the period of the flow, $f_i.length$ 
the packet length of the flow in one period, and $f_i.delay$ 
the e2e delay boundary required by the flow.

For each flow, we assume there is only one frame to be delivered 
within one period, and the route is determined in advance: 

\begin{equation}
\forall f_i\in F:p_i = [l_1,l_2,\ldots,l_{h_i}],l_j\in L,
\end{equation}
where $h_i$ is the number of hops for $f_i$. Table \ref{table:notations} 
summarizes the main notations/symbols used in this paper.

\begin{figure}[!t]
\centering
\includegraphics[width=3.1in]{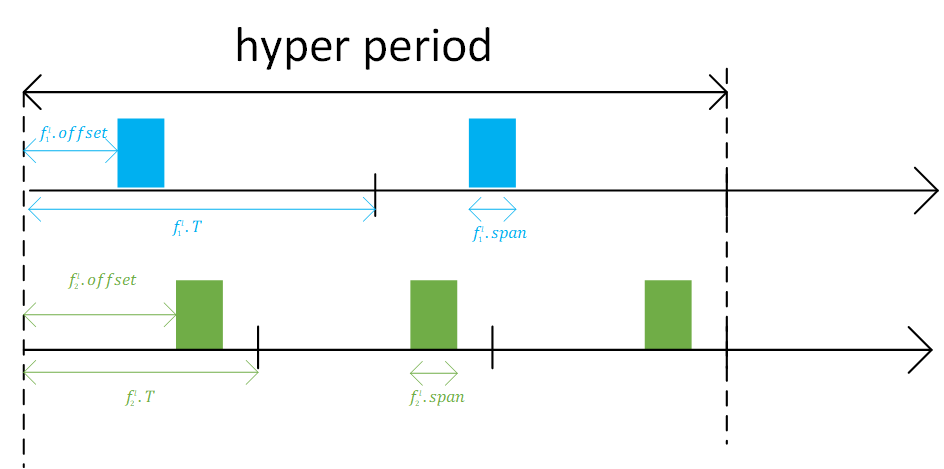}
\caption{Scheduling example in TSN.}
\label{fig:TSNResource}
\end{figure}

\begin{table}[]
\caption{Notations used in this work.}
\begin{tabular}{|c|c|}
\hline
\textbf{Symbol}                                                                       & \textbf{Description}                                                                                                    \\ \hline
$G$                                                                                   & the wired TSN network                                                                                                   \\ \hline
$V$                                                                                   & the set of nodes in TSN                                                                                                 \\ \hline
$E$                                                                                   & the set of wired links in TSN                                                                                           \\ \hline
$F$                                                                         & \begin{tabular}[c]{@{}c@{}}the set of all flows\end{tabular}                        \\ \hline
$L, L^{TSN}, L^{5GS}$                                                                  & \begin{tabular}[c]{@{}c@{}}the set of dataflow links\\ of the whole network, TSN, and 5GS\end{tabular}                   \\ \hline
$v_i\in V$                                                                         & \begin{tabular}[c]{@{}c@{}}the node of TSN\end{tabular}                        \\ \hline
$(v_1, v_2)\in E$                                                                     & \begin{tabular}[c]{@{}c@{}}the physical link\\ between nodes $v_1$ and $v_2$\end{tabular}                               \\ \hline
$[v_1, v_2], [v_2, v_1]\in L^{TSN}$                                                              & \begin{tabular}[c]{@{}c@{}}the dataflow links\\ between nodes $v_1$ and $v_2$\end{tabular}                              \\ \hline
$f_i\in F$                                                                         & \begin{tabular}[c]{@{}c@{}}one flow of the set F\end{tabular}                        \\ \hline
$p_i$                                                                                 & the route of flow $f_i$                                                                                                 \\ \hline
$h_i$                                                                                 & \begin{tabular}[c]{@{}c@{}}the number of hops\\ of the route $p_i$\end{tabular}                                 \\ \hline
$T_i$                                                                                 & \begin{tabular}[c]{@{}c@{}}the period of transmission\\ opportunities reserved in TSN for $f_i$\end{tabular}                          \\ \hline
\begin{tabular}[c]{@{}c@{}}$f_i.period$, \\ $f_i.length, f_i.delay$\end{tabular}      & \begin{tabular}[c]{@{}c@{}}the period, length, \\ and delay requirement of $f_i$\end{tabular}                         \\ \hline
$f_i^l$                                                                               & \begin{tabular}[c]{@{}c@{}}the potential transmission \\ instance of $f_i$ on dataflow link $l$\end{tabular}            \\ \hline
\begin{tabular}[c]{@{}c@{}}$f_i^{l}.T$, \\ $f_i^l.offset, f_i^l.span$\end{tabular} & \begin{tabular}[c]{@{}c@{}}the transmission period, start offset, \\ and transmission delay of the $f_i^l$\end{tabular} \\ \hline
$\mathcal{F}$                                                                         & \begin{tabular}[c]{@{}c@{}}the whole set of RBs in 5GS\end{tabular}            \\ \hline
$k,k_{max}$                                                                           & \begin{tabular}[c]{@{}c@{}}the $k^{th}$ RB and the \\ whole number of RBs\end{tabular}                \\ \hline
$x^l_{i,k}$                                                                           & \begin{tabular}[c]{@{}c@{}}whether the $k^{th}$ RB of \\ dataflow link $l$ is assigned to $f_i$\end{tabular}   \\ \hline
$y^l_k$                                                                               & \begin{tabular}[c]{@{}c@{}}whether the $k^{th}$ RB of\\ dataflow link $l$ is assigned to any flow\end{tabular} \\ \hline
$minP$                                                                                & the minimum period supported by TSN                                                                                     \\ \hline
$T_{TSN}$                                                                                & the TSN hyper period                                                                                     \\ \hline
$T_{5GS}$                                                                                & the 5GS hyper period                                                                                     \\ \hline
\end{tabular}
\label{table:notations}
\end{table}

\subsection{Scheduling Model}
We formally define the transmission opportunities reserved in TSN 
and 5GS as potential transmission instances.\footnote{Since each 
transmission opportunity is allocated to an actually transmitted 
packet in the traditional traffic scheduling model of TSN, 
each transmission opportunity is also called a {\em transmission instance} \cite{zhang2022survey}. However, in the AAM context, 
not every transmission opportunity corresponds to an actually 
transmitted packet. So, we extend the concept of transmission 
instances to potential transmission instances.}
Let $f^{l}_i$ denote the potential transmission instance of flow 
$f_i$ on the dataflow link $l$.

\subsubsection{Scheduling model in TSN}
In TSN, the potential transmission instance $f^{l}_i$ is fully 
determined by the following triple:
\begin{align}
    &\forall f_i\in F,\forall l\in p_i\cap L^{TSN}:\nonumber\\
    &f_i^{l} = \{f_i^{l}.T,f_i^{l}.offset,f_i^{l}.span\}
\end{align}
where $f_i^l.T$ denotes the period of the transmission 
opportunities reserved for $f_i$ on the dataflow link $l$, 
while $f_i^l.offset$ and $f_i^l.span$ denote the offset and 
the transmission duration. 
Fig.~\ref{fig:TSNResource} shows an example.

For a single flow, $f_i^l.T$ is the same along its route 
(denoted as $T_i$) and $f_i^l.span$ is the ratio of the frame 
length to the link rate which is known {\em a priori}, 
and the pattern repeats in every cycle. For multiple flows, 
the pattern similarly repeats in every "hyper period". 
TSN's hyper period is defined as the least common multiple of 
the periods of every flow's potential transmission instances:
\begin{gather}
    T_{TSN}=LCM(\{T_i|f_i\in F\}).
\end{gather}
As shown in Fig.~\ref{fig:TSNResource}, the reference zero points of 
all potential transmission instances of all flows are the same, and 
in one hyper period, the potential transmission instance of $f_1$ on 
dataflow link $l$ appears twice while that of $f_2$ appears thrice. 

The values of $f_i^{l}.T$ and $f_i^{l}.offset$ for all flows 
$f_i\in F$ on TSN dataflow links of its route 
$l\in p_i\cap L^{TSN}$(except for the last dataflow link) are design variables.

\subsubsection{Scheduling model in 5GS}

\begin{figure}[!t]
\centering
\includegraphics[width=3.1in]{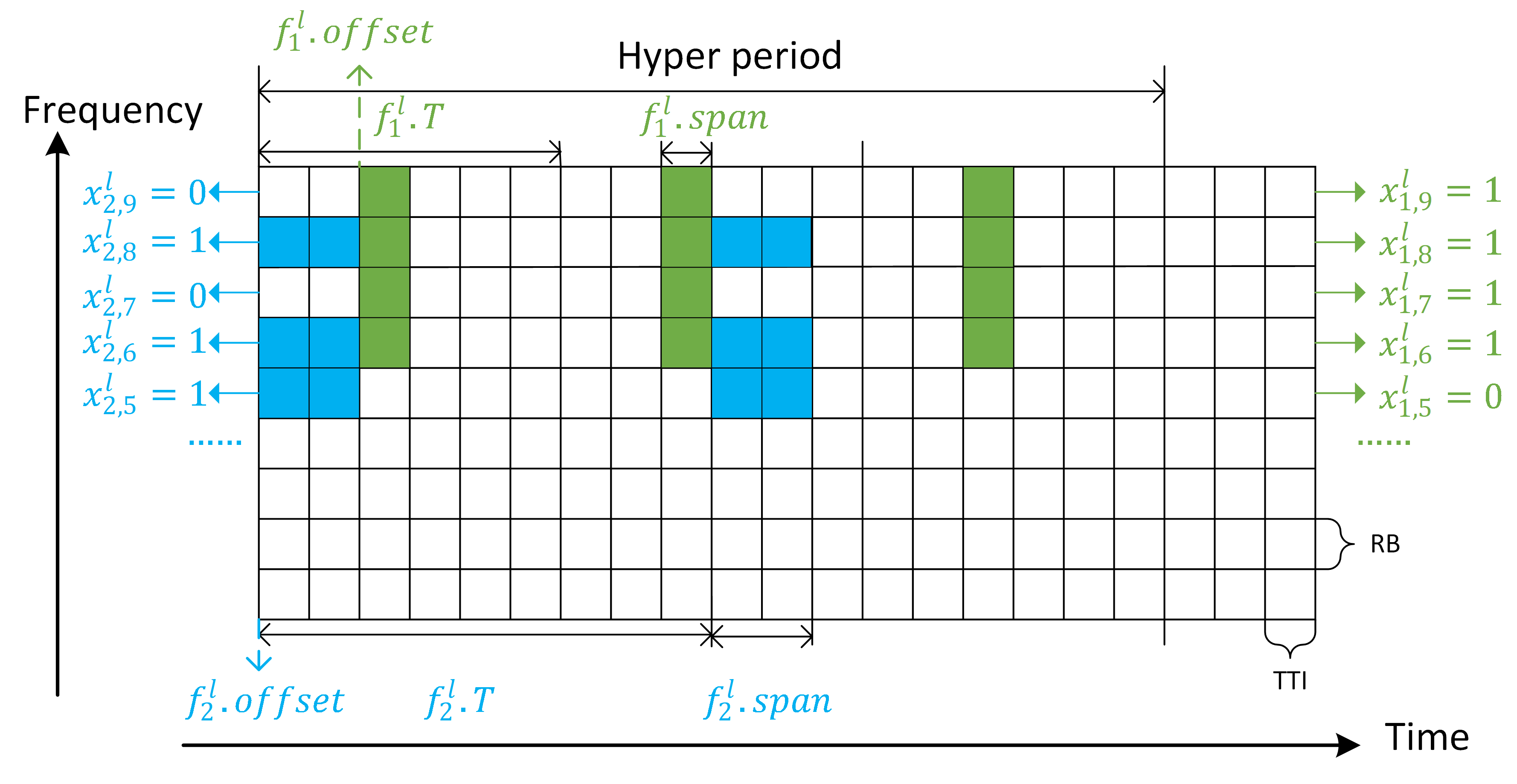}
\caption{Scheduling example in 5GS.}
\label{fig:5GResource}
\end{figure}

As shown in Fig.~\ref{fig:5GResource}, the resources in 5GS can be 
split into two dimensions: transmission time intervals (TTI) in 
the time domain and the resource blocks (RBs) in the frequency domain.

Assuming there can be at most $k_{max}$ RBs being allocated to 
these flows, the set of RBs is denoted as $\mathcal{F}=\{1, 2, 
\ldots, k_{max}\}$. In 5GS, we use the SPS that reserves periodic 
resources for transmission \cite{Jiang2007PrincipleAP}. 
To reduce the scheduling complexity, one potential transmission 
instance in 5GS can only occupy consecutive TTIs but we do not 
impose any constraint on the frequency domain. For example, as shown 
in Fig.~\ref{fig:5GResource}, the potential transmission instances 
of $f_2$ occupy consecutive TTIs but non-consecutive RBs. 

The potential transmission instance in 5GS can thus be identified 
by the following variables:
\begin{align}
    &\forall f_i\in F,\forall l\in p_i\cap L^{5GS}: \nonumber\\
    &f_i^{l} = \{f_i^{l}.T,f_i^{l}.offset,f_i^{l}.span,x^{l}_i\}\\ 
    &x^{l}_i = \{x^{l}_{i,k}|k\in\mathcal{F}\}
\end{align}
where $f_i^{l}.T$ is the flow's period and $x^{l}_{i,k}$ is 
a binary variable equal to 1 (0) if flow $f_i$ is (not) allocated 
to RB $k$ on dataflow link $l$. The variables $x_{i,k}^l$ allow us 
to schedule the RBs assigned to the flow while the variables 
$f_i^l.span$ are used to schedule the consecutive TTIs.

The concept of the hyper period also applies to 5GS:
\begin{gather}
    T_{5GS}=LCM(\{f_i.period|f_i\in F\}).
\end{gather}

The values of $f^l_i.offset$, $f^l_i.span$ and $x^l_{i,k}$ for all 
flows $f_i\in F$ on 5G dataflow links of its route 
$l\in p_i\cap L^{5GS}$ and each RB $k\in\mathcal{F}$ are design 
parameters.

\subsection{Network Constraints}
\label{uplink-constraints}
We now derive the constraints of the ATSM which accounts for both 5GS 
and TSN and schedules flows from an e2e perspective. The constraints 
introduced in \ref{Frame constraints}, \ref{Transmission order 
constraints}, \ref{TDMA constraints}, and \ref{Frame isolation 
constraints} are borrowed from the mature scheduling model in 
TSN \cite{Raagaard2018OptimizationAF,Ginthr2020EndtoendOJ}, but are 
modified, if needed, for the AAM.

\subsubsection{Transmission-opportunity constraints}
Every packet in 5GS cannot start its transmission except for the 
start of TTI and the transmission time of a packet must be an
integer multiple of TTI:
\begin{align}
&\forall l\in L^{5GS},\forall f_i\in F:\nonumber\\
&\exists f_i^l\Rightarrow \nonumber\\
&f_i^l.offset = c_i^l\times TTI\\
&f_i^{l}.span = d_i^l\times TTI\\
&f_i^l.offset+f_i^{l}.span\leq f_i.period\\
&c_i^l\in \{0,1,\ldots,\lfloor\frac{f_i.period}{TTI}
\rfloor-1\}\nonumber\\
&d_i^l\in\{1,2,\ldots,\lfloor\frac{f_i.period}{TTI}\rfloor\},
\nonumber
\end{align}
where $c_i^l$ denotes the TTI flow $f_i$ can start transmission in 
the 5G dataflow link $l$ and $d_i^l$ is the number of consecutive 
TTIs occupied by the flow on the allocated RBs.

\subsubsection{RB constraints}  
We use binary variables $y^{l}_{k}$ to indicate whether the $k^{th}$ 
RB of 5G dataflow link $l$ is scheduled to any flow. Specifically, 
$y^l_k=1 (0)$ if it is (not) assigned to any flow. 
So, we derive the following constraints:
\begin{align}
&\forall l\in L^{5GS},\forall k\in \mathcal{F}:\nonumber\\
&\sum\limits_{f_i\in F}x^{l}_{i,k} \leq y^l_{k} \times M,
\end{align}
where $M$ is a (theoretically) infinitely large constant. 
The constraints guarantee that if $y_k^l=0$, all $x^l_{i,k}$ 
for all flows should be 0.

\subsubsection{Resource constraints}
The resources allocated to a flow should be sufficient to transmit 
its packets. Let $R^{l}_{i,k}$ denote the number of bytes that can be 
transmitted by the $k^{th}$ RB on the dataflow link $l$ for the flow 
$f_i$  which can be obtained from the simulator, then we have:
\begin{align}
&\forall l\in L^{5GS},\forall f_i\in F:\nonumber\\
&\exists f_i^l \Rightarrow \nonumber\\
&\sum\limits_{k=1}^{k_{max}}x^{l}_{i,k}\times R^{l}_{i,k}\times d_i^l\geq f_i.length\label{eq:resource1}\\
&\sum\limits_{k=1}^{k_{max}}x^{l}_{i,k}\times R^{l}_{i,k}\times (d_i^l-1)< f_i.length\label{eq:resource2}\\
&1 \leq d_i^l\leq \lfloor\frac{f_i.period}{TTI}.\rfloor\nonumber
\end{align}

\subsubsection{OFDMA constraints}
Any two flows cannot be assigned to the same TTI on the same RB just 
as shown for $f_1$ and $f_2$ in Fig.~\ref{fig:5GResource}:
\begin{align}
& \forall l\in L^{5GS},\forall f_i,f_j\in F,\forall k\in\mathcal{F},\nonumber\\
&\forall\alpha\in[0..(\frac{T_{5GS}}{f_i.period}-1)],\nonumber\\
&\forall\beta\in[0..(\frac{T_{5GS}}{f_j.period}-1)]:\nonumber\\
&((f_i\neq f_j)\land\exists f^l_i\land\exists f^l_j)\Rightarrow\nonumber\\
&((\alpha\times f_i.period)+f_i^l.offset+M\times (2-x^l_{i,k}-x^l_{j,k})\geq\nonumber\\
&(\beta\times f_j.period)+f_j^l.offset+f_j^l.span)\vee\nonumber\\
&((\beta\times f_j.period)+f_j^l.offset+M\times(2-x^l_{i,k}-x^l_{j,k})\geq\nonumber\\
&(\alpha\times f_i.period)+f_i^l.offset+f_i^l.span).\label{eq:OFDMA}
\end{align}

\subsubsection{Window constraints}
Potential transmission instances have the same period 
along the route of the flow, and the period should not be 
larger than that of the flow. Therefore,
\begin{align}
&\forall l\in L^{TSN},\forall f_i\in F:\nonumber\\
&(\exists f_i^{l}\land (l\neq l_{h_i}))\Rightarrow \nonumber \\
&f^l_i.T = T_i\\
&f^l_i.T \leq f_i.period.
\end{align}

\subsubsection{Frame constraints}\label{Frame constraints}
The offset should be positive, and for simplicity 
a potential transmission instance should finish the transmission
within its period:
\begin{align}
&\forall l\in L^{TSN},\forall f_i\in F:\nonumber\\
&(\exists f_i^{l}\land (l\neq l_{h_i}))\Rightarrow \nonumber \\
&f^l_i.offset\geq 0\\
&f^l_i.offset + f^l_i.span \leq T_i.
\end{align}

\subsubsection{Transmission order constraints}
\label{Transmission order constraints}
The transmission opportunity on the next wired dataflow link $l_b$ 
should send the packet after it is fully received over the 
preceding wired dataflow link $l_a$:
\begin{align}
&\forall l_a,l_b\in L^{TSN},\forall f_i\in F:\nonumber\\
&(\exists f_i^{l_{a}}\land\exists f_i^{l_{b}}\land 
isNextHop(l_b,l_a,f_i)\land (l_b\neq l_{h_i}))\Rightarrow \nonumber \\
&f_i^{l_{b}}.offset \geq f_i^{l_{a}}.offset + f_i^{l_{a}}.span 
+ ldelay^{l_a},
\end{align}
where $ldelay^{l_a}$ is the propagation delay of the 
dataflow link $l_a$ and function $isNextHop(l_b,l_a,f_i)$ returns 
true if $l_b$ is the next dataflow link of $l_a$ on the route of 
flow $f_i$ else false. 

\subsubsection{TDMA constraints}
\label{TDMA constraints}
Wired links can only transmit one frame at a time, and hence  
potential transmission instances reserved for different 
flows should not overlap:
\begin{align}
&\forall l\in L^{TSN},\forall f_i,f_j\in F,\nonumber\\
&\forall\alpha\in[0..(\frac{T_{TSN}}{T_i}-1)],\nonumber\\
&\forall\beta\in[0..(\frac{T_{TSN}}{T_j}-1)]:\nonumber\\
&((f_i\neq f_j)\land\exists f^l_i\land\exists f^l_j\land (l\neq l_{h_i})\land (l\neq l_{h_j}))\Rightarrow\nonumber\\
&((\alpha\times T_i) + f_i^{l}.offset\geq\nonumber\\
&(\beta\times T_j) + f_j^{l}.offset+f_j^l.span)\vee\nonumber\\
&((\beta\times T_j)+f_j^l.offset\geq\nonumber\\
&(\alpha\times T_i)+f_i^l.offset+f_i^l.span).\label{eq:tdma}
\end{align}
Note that the $T_{TSN}$ contains the scheduling parameters while 
it is fixed in the traditional scheduling of TSN. We will discuss 
how to linearize these constraints in Section \ref{sec:linearization}.

\subsubsection{Frame isolation constraints}
\label{Frame isolation constraints}

If frames of different flows arrive at the same TSN switch at the 
same time and have the same output port, the order that the frames 
in the queue is not certain hence introducing non-determinism 
\cite{Raagaard2018OptimizationAF}. So, only frames from the same 
flow can be stored in the queue:
\begin{align}
&\forall l_a,l_b,l_c\in L^{TSN},\forall f_i,f_j\in F,\nonumber\\
&\forall\alpha\in[0..(\frac{T_{TSN}}{T_i}-1)],\nonumber\\
&\forall\beta\in[0..(\frac{T_{TSN}}{T_j}-1)]:\nonumber\\
&((f_i\neq f_j)\land\nonumber\\
&\exists f_i^{l_a}\land\exists f_i^{l_c}\land 
\exists f_j^{l_b}\land\exists f_j^{l_c}\land\nonumber\\ 
&isNextHop(l_c,l_a,f_i)\land \nonumber\\
&isNextHop(l_c,l_b,f_j)\land \nonumber\\
&(l_c\neq l_{h_i})\land (l_c\neq l_{h_j}))\Rightarrow\nonumber\\
&(((\alpha\times T_i)+f_i^{l_c}.offset\leq \nonumber\\
&(\beta\times T_j)+f_j^{l_b}.offset+ldelay^{l_b})\vee \nonumber\\
&((\beta\times T_j)+f_j^{l_c}.offset\leq \nonumber\\
&(\alpha\times T_i)+f_i^{l_a}.offset+ldelay^{l_a})
\label{eq:Frame isolation}
\end{align}

If $f_i$ and $f_j$ come from the same input port, $l_a$ and $l_b$ 
can be the same dataflow link.

\subsubsection{e2e delay constraints}
The e2e delay of a flow consists of 5GS and TSN parts.
For the 5GS part, the delay is actually $f^{h_1}_i.span+T_{proc,gNB}$, 
where $T_{proc,gNB}$ is the (constant) processing delay of BS to 
receive packets. 
For TSN part, the delay is always fixed due to the AAM. 
It can be calculated using Eq.~\eqref{eq:delayTSN} in 
Section \ref{sec:basicIdeasofAAM}.
Thus, the e2e delay can be calculated by Eq.~\eqref{eq:e2eDelay}, 
and its elements that make up the formula are calculated as:

\begin{gather}
    D^{5GS}_i = f^{h_1}_i.span+T_{proc,gNB}
\end{gather}
\begin{gather}
        D^{TSN}_{i,(1)} = (f_i^{l_{h_i-1}}.offset+f_i^{h_{i-1}}.span+ldelay^{l_{h_{i-1}}})\nonumber\\
        - f_i^{l_2}.offset
\end{gather}
\begin{gather}
        D^{TSN}_{i,(2)} = f_i^{h_{i}}.span+ldelay^{l_{h_{i}}}.
\end{gather}

The e2e delay constraints can thus be described as:
\begin{align}
&\forall f_i\in F:\nonumber\\
&D^{e2e}_i \leq f_i.delay. 
\end{align}

\subsection{Objective Function}
In addition to the TT flows, many other types of flows (e.g.,
best-effort (BE) flows) may exist in an industrial setting. 
We need to reduce the resources used by TT flows and leave as 
much resource as possible for the BE flows.

For the 5GS part, we try to minimize the number of RBs assigned to 
TT flows, which is inspired by \cite{Karadag2019QoSConstrainedSS}
(we omit the superscript $l$ for readability as there is only one 
dataflow link in 5GS on the uplink):
\begin{gather}
\mathop{min}\sum\limits_{k\in\mathcal{F}}\frac{y_{k}}{|\mathcal{F}|}.
\end{gather}

For the TSN part, as we adopt the AAM, over-provisioned transmission 
opportunities will lead to unused resources. Thus, in TSN, we want 
to maximize the periods of the transmission opportunities 
reserved for the flows to reduce waste of resources:
\begin{gather}
    \mathop{max} \sum \limits_{f_i\in F} \frac{T_i}{f_i.period|F|}. 
    \label{eq:optim-tsn}
\end{gather}

To obtain the final objective function, we multiply 
Eq.~\eqref{eq:optim-tsn} by -1 to change the optimization direction, 
and then use a weighting factor $\gamma$ to balance 
the two optimization objectives:
\begin{gather}
    \mathop{min}  [\gamma \sum\limits_{k\in\mathcal{F}}\frac{y_{k}}{|\mathcal{F}|}-(1-\gamma)\sum \limits_{f_i\in F} \frac{T_i}{f_i.period\cdot |F|}],
\end{gather}
where $\gamma\in [0,1]$ indicates which network we want to 
optimize more.

\subsection{Problem Linearization}
\label{sec:linearization}
The ATSM proposed above can be transformed into an Integer
Linear Programming problem. The "logical or" in Eqs.~\eqref{eq:OFDMA}, \eqref{eq:tdma} and \eqref{eq:Frame isolation} 
and the product of binary variable and bounded variable in 
Eqs.~\eqref{eq:resource1} and \eqref{eq:resource2} can be linearized by 
the commonly used linearization methods \cite{Boyd2004ConvexO}. 
But there still remains a difficulty: the number of constraints in 
Eqs.~\eqref{eq:tdma} and \eqref{eq:Frame isolation} cannot be 
determined as the number is related to the $T_i$ and  $T_j$ 
which are decision variables.



To overcome this difficulty, we adopt the following strategy: 
\begin{itemize}
  \item The minimum resource period can be supported by TSN is 
    denoted as $minP$ whose unit is $\mu s$.
  \item For any flow $f_i$ whose period is $f_i.period$, the 
  available resource period for it can only be selected from the 
  list $list_i=[minP,2\times minP,4\times minP,\ldots,T_{i,max}]$ 
  whose length is $s_i$ and $T_{i,max}$ satisfies: 
  \begin{gather}
    T_{i,max}=minP\cdot 2^{s_i-1}\\
    minP\times 2^{s_i-1} \leq f_i.period\\
    minP\times 2^{s_i} > f_i.period.
  \end{gather}
  \item For any flow $f_i$ we introduce $s_i$ binary variables 
  $\{b_{i,j}|j=0,1,\ldots,s_i-1\}$, and we let 
  $T_i=\sum\limits_{j=0}^{j=s_i-1}list_i[j]\times b_{i,j}$. 
  Furthermore, we introduce new constraints to make only one 
  element of the list selected:
  \begin{gather}
    \forall f_i\in F:\nonumber\\
    \sum\limits_{j=0}^{j=s_i-1}b_{i,j}=1.
  \end{gather}
  
\end{itemize}

This way, the hyper period of the 
set $\mathbf{T}_{max}=\{T_{i,max}|f_i\in F\}$ 
is multiple of the hyper period of the set $\mathbf T=\{T_i|f_i
\in F\}$. Then, we can expand the check range from 
$[0,\frac{T_{TSN}}{T_i}-1]$ to $[0,\frac{LCM(\mathbf {T}_{max})}
{minP}-1]$ in Eqs.~\eqref{eq:resource1} and \eqref{eq:resource2}.

The optimization problem is transformed into an Integer Linear 
Programming problem. We use Gurobi \cite{gurobi} to solve the problem.

\section{performance evaluation}
\label{PerformanceEvaluation}

\begin{figure}[!t]
\centering
\includegraphics[width=3.5in]{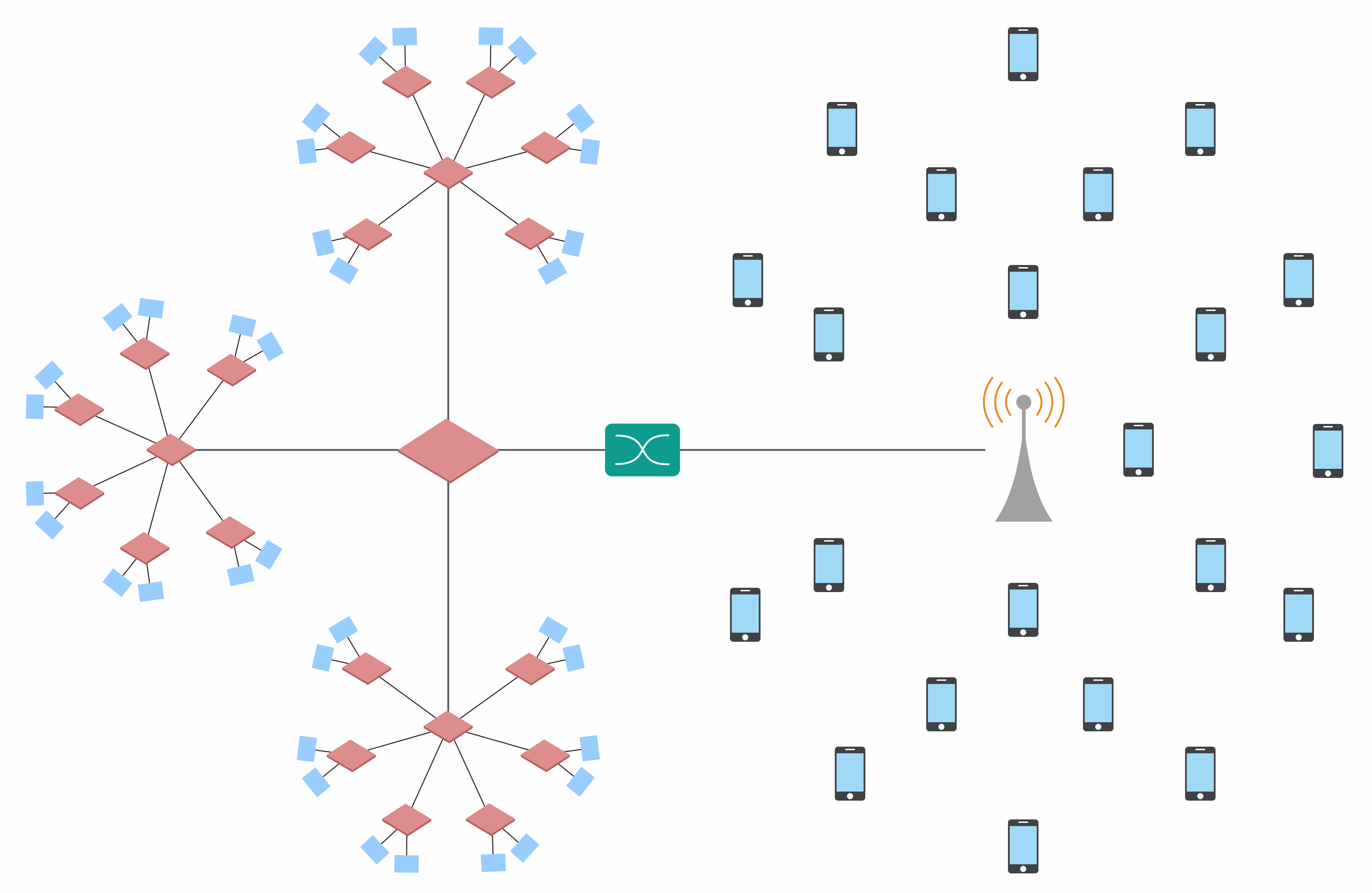}
\caption{Simulation network topology consisting of one 5GS 
and a TSN network.}
\label{fig:simulationTopology}
\end{figure}

\begin{table}[]
\caption{Test cases}
\centering
\begin{tabular}{ccccc}
\hline
\multicolumn{1}{l}{\textbf{Flow type}} & \multicolumn{1}{l}{\textbf{Period}} & \multicolumn{1}{l}{\textbf{Size}} & \multicolumn{1}{l}{\textbf{Delay}} & \multicolumn{1}{l}{\textbf{Amount}} \\ \hline
\uppercase\expandafter{\romannumeral1}                                      & $0.5ms$                               & $96B$                               & $0.5ms$                              & 5                                   \\
\uppercase\expandafter{\romannumeral2}                                      & $1ms$                                 & $128B$                              & $1ms$                                & 5                                   \\
\uppercase\expandafter{\romannumeral3}                                      & $2ms$                                 & $256B$                              & $2ms$                                & $10$                                  \\ \hline
\end{tabular}
\label{table:flowsSettings}
\end{table}
\begin{table}[]
\caption{Simulation Parameters}
\begin{tabular}{lll}
\hline
\textbf{Type}             & \textbf{Parameter}              & \textbf{Value}               \\ \hline
\multirow{6}{*}{Wireless} & Subcarriers spacing             & 120kHz                       \\
                          & Mini-slot duration (TTI)        & 7 OFDM symbols               \\
                          & $k_{max}$                         & 10                           \\
                          & $T_{proc,gNB}$ & 1 TTI                        \\
                          & Carrier Frequency               & 3.8 GHz                      \\
                          & Channel Model                   & Urban Macrocell according to\cite{3gpp.38.901} \\ \hline
\multirow{2}{*}{Wired}    & Datarate                        & 100 Mbps                       \\
                          & Propagation delay               & $1us$                         \\ 
                          & minP                            & $100us$                       \\\hline
\end{tabular}
\label{table:parameters}
\end{table}

We evaluate the performance of AAM and ATSM. 
All simulations are conducted with the discrete-event simulator 
OMNET++\cite{OMNET} with INET\cite{INET} and 
Simu5G\cite{9211504} framework.

\subsection{Simulation Setup}
The topology of the simulation network is shown in 
Fig.~\ref{fig:simulationTopology}, which 
has referenced the medium-sized topology in \cite{Raagaard2018OptimizationAF} and is extended with a 5GS, 
where blue squares represent ESs and red diamonds represent 
TSN switches. There are 36 ESs in TSN and 20 UEs uniformly 
distributed around the base station in 5GS. There are a total of 
20 flows, consisting of three types, and the specific configuration 
is shown in Table \ref{table:flowsSettings}, which has referenced 
\cite{industrial2019time}. The remaining main simulation 
parameters are summarized in Table \ref{table:parameters}.

\subsection{Evaluation of AAM}

In the first set of experiments, we evaluate the performance of 
AAM compared to TAM in the presence of the 5G transmission jitter 
and the clock skew between the two systems. To distinguish between 
different experiments, we name the experiment conducted by the ATSM 
introduced in Section \ref{asynchronous traffic scheduling model} 
as the AAM-Scheduling. The baseline is called the 
{\em TAM-Scheduling}, 
which is done with a scheduling model called the 
{\em synchronous traffic scheduling model} (STSM). 
The STSM is modified from the ATSM by not applying the AAM, i.e., 
the period of the potential transmission instances equals the 
flow period, the packets have to arrive at the GW before the start of 
the transmission opportunity prepared for it, and no holding process 
is adopted as in \cite{Ginthr2020EndtoendOJ}.

As shown in Eqs.~\eqref{eq:CE} and \eqref{eq:CV}, we use the coefficient 
of expansion (CE) and coefficient of variation (CV) of the e2e delay 
to evaluate the ratios of e2e delay enlargement and 
e2e jitter, respectively, 
which are defined respecively as the ratio of the mean actual 
e2e delay to the scheduled delay and that of the standard 
deviation of the e2e delay to the mean actual e2e delay:
\begin{align}
    &\forall f_i \in F:\nonumber\\ 
    &CE_i = \frac{\bar{D}^{e2e}_{i,Actual}}{D^{e2e}_{i,Scheduled}}\label{eq:CE}\\
    &CV_i = \frac{\sigma(D^{e2e}_i)}{\bar{D}^{e2e}_{i,Actual}},\label{eq:CV}
\end{align}
where $D^{e2e}_{i,Scheduled}$ is the e2e delay calculated by the
ATSM or the STSM, which both assume no transmission jitter in 5GS, 
$\bar{D}^{e2e}_{i,Actual}$ is the mean e2e delay the packets actually 
experience, and $\sigma (\cdot)$ is the standard deviation calculated 
by accounting for the e2e delay of each data packet.

Then, the mean coefficient of expansion (MCE) and mean coefficient 
of variation (MCV) are calculated to evaluate the ratios of
e2e delay enlargement and jitter of the whole system:
\begin{align}
    MCE &= \sum\limits_{f_i\in F}\frac{CE_i}{|F|}\\
    MCV &= \sum\limits_{f_i\in F}\frac{CV_i}{|F|}.
\end{align}

\subsubsection{Performance with 5G transmission jitter}

\begin{figure}[!t]
\centering
\includegraphics[width=3.5in]{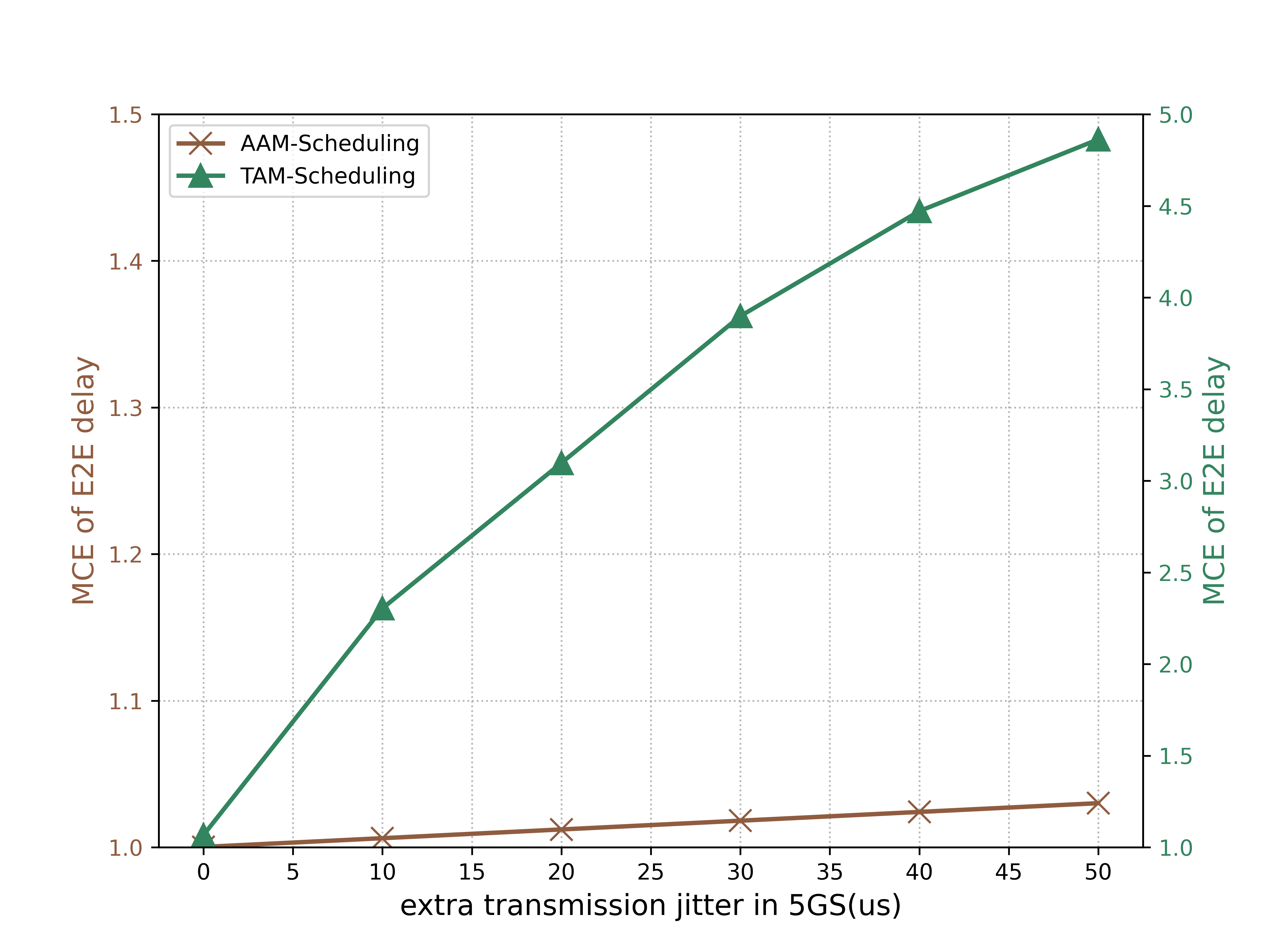}
\caption{The mean coefficient of expansion (MCE) of e2e delay 
varies with extra 5G transmission jitter.}
\label{exp-AAM-1}
\end{figure}

\begin{figure}[!t]
\centering
\includegraphics[width=3.5in]{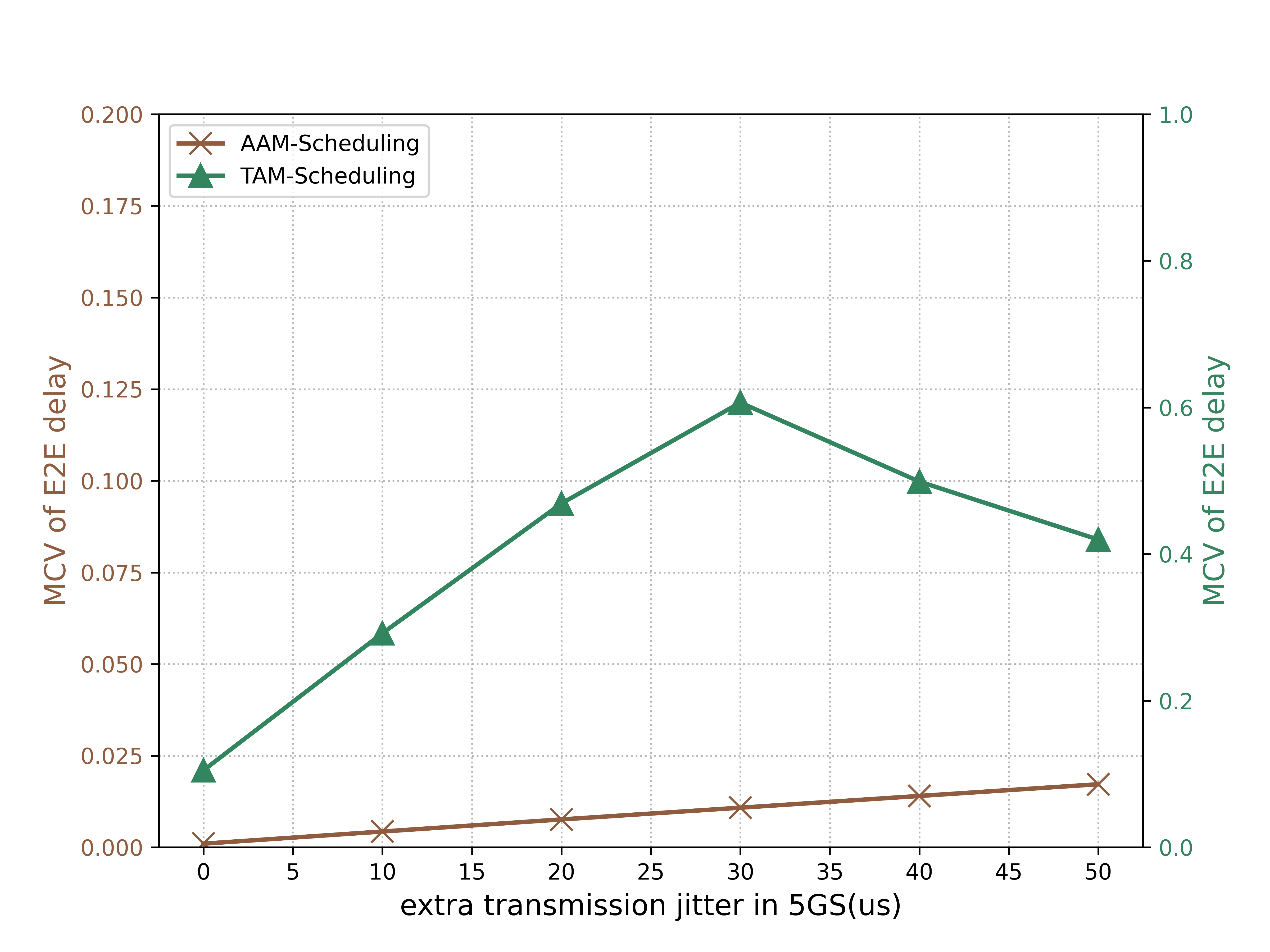}
\caption{The mean coefficient of variance (MCV) of e2e delay 
varies with extra 5G transmission jitter.}
\label{exp-AAM-2}
\end{figure}

\begin{figure}[!t]
\centering
\includegraphics[width=3.5in]{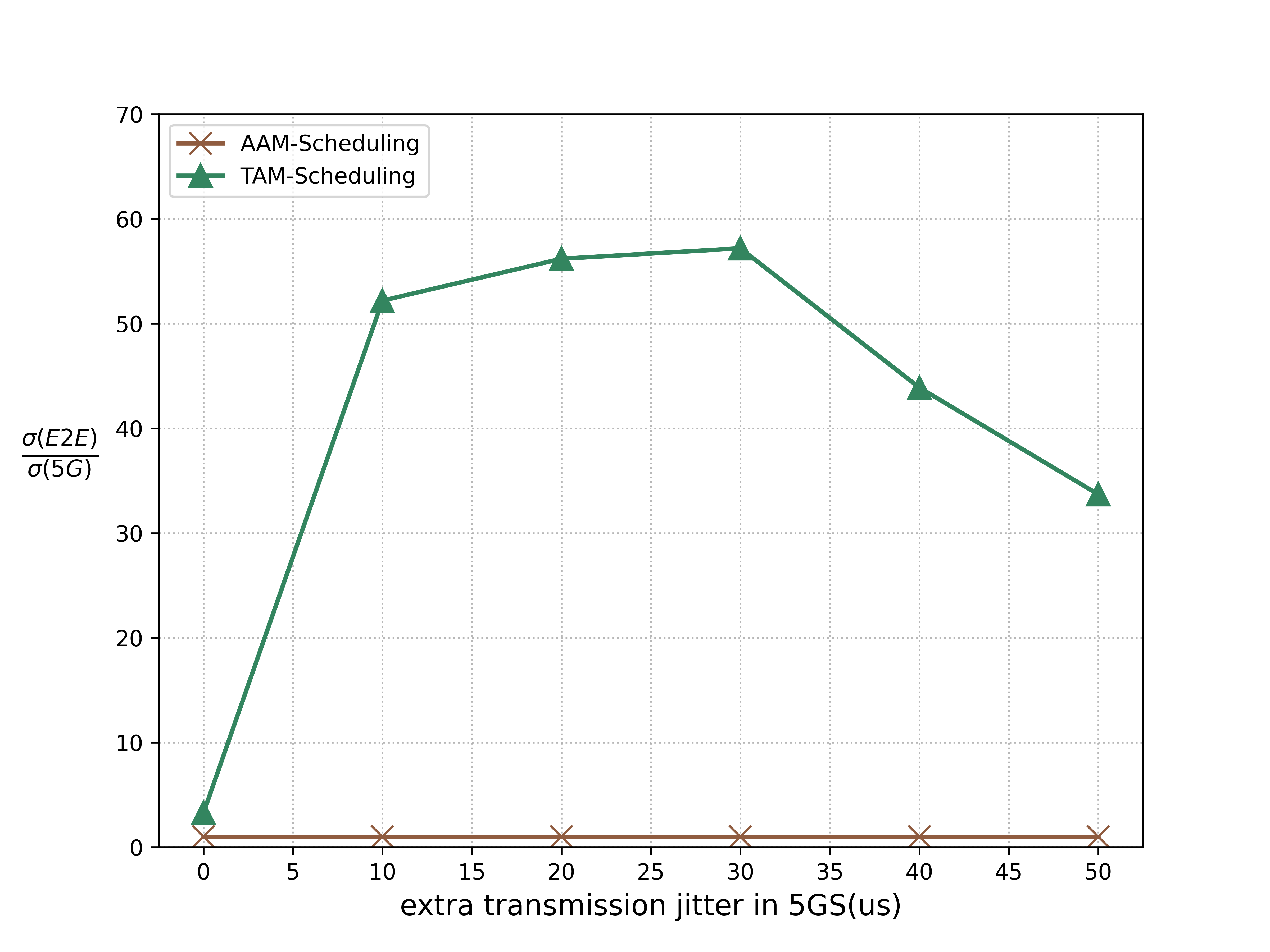}
\caption{The ratio of the standard deviation of e2e delay and to the 
standard deviation of 5GS delay varies with extra 5G transmission jitter.}
\label{exp-AAM-3}
\end{figure}

We first assume 5GS and TSN are synchronized and introduce an 
extra transmission jitter in 5GS in addition to the original 5G 
transmission jitter. Specifically, We vary the extra transmission 
jitter from $0us$ to $50us$ with step $10us$ by introducing a 
random extra delay in 5GS for each packet, and the extra delay 
follows a uniform distribution. For example, if the extra 
transmission jitter equals $10us$, the introduced extra delay for
a packet in 5GS is sampled from the uniform distribution 
$U(0us,10us)$. The results are plotted in Figs.~\ref{exp-AAM-1}, \ref{exp-AAM-2} and \ref{exp-AAM-3}.

In Fig.~\ref{exp-AAM-1}, the MCE of the e2e delay of the 
TAM-Scheduling increases rapidly as the extra transmission 
jitter increases, since more packets will miss their corresponding 
transmission opportunities and a miss will incur an additional 
delay equal to the flow period as shown in Fig.~\ref{fig:AAM-2}. 
However, the MCE of the AAM-Scheduling rather increases slowly 
compared to that of the TAM-Scheduling because a miss of transmission 
opportunity has no impact on the constant delay in TSN as shown in 
Fig.~\ref{fig:AAM-4} and Eq.~\eqref{eq:delayTSN}, and the e2e delay 
increases with the delay in 5GS which is affected by the 
thus-introduced mean extra transmission delay. Thus, the delay 
enlargement is significantly mitigated by the AAM, 
which is consistent with Fig.~\ref{fig:AAM-4}. 

Figs.~\ref{exp-AAM-2} and \ref{exp-AAM-3} show the MCV and ratio of 
the standard deviation of e2e delay to the standard deviation of 
5GS delay for the TAM-Scheduling and the AAM-Scheduling. 
In these two figures, the curves of the TAM-Scheduling first 
increase rapidly, because more packets will miss their transmission 
opportunities as the extra transmission jitter increases, and then
they will decrease after the peak because more packets are 
transmitted in the next transmission opportunities in their next 
flow periods, thus decreasing the e2e jitter. However, the MCV of 
the AAM-scheduling increases rather slowly compared to that of the 
TAM-Scheduling, and the ratio of the standard deviation of the e2e 
delay to the standard deviation of 5GS delay is always 1 as shown 
in Fig.~\ref{exp-AAM-3}. This is because the transmission delay in 
TSN is constant and the e2e jitter is only contributed by the 5G 
transmission jitter. From the above two figures, one can see that 
the jitter enlargement is mitigated significantly by the AAM as well.

However, the AAM provides such excellent transmission performance
for TT flows at the cost of over-provisioning the transmission 
opportunities, wasting more resources. Even though the traffic 
load is the same in the two schedules, TSN resource usage by 
the TT flows in the AAM-Scheduling is 44.24\% while the 
TAM-Scheduling only uses 23.04\%, where TSN resource usage is
defined as the ratio of the sum of the time duration when 
the gateway’s output port is open for TT flows to the
total simulation time.

\subsubsection{Performance under clock skew}
\label{uplink-clockerror}

\begin{figure}[!t]
\centering
\includegraphics[width=3.5in]{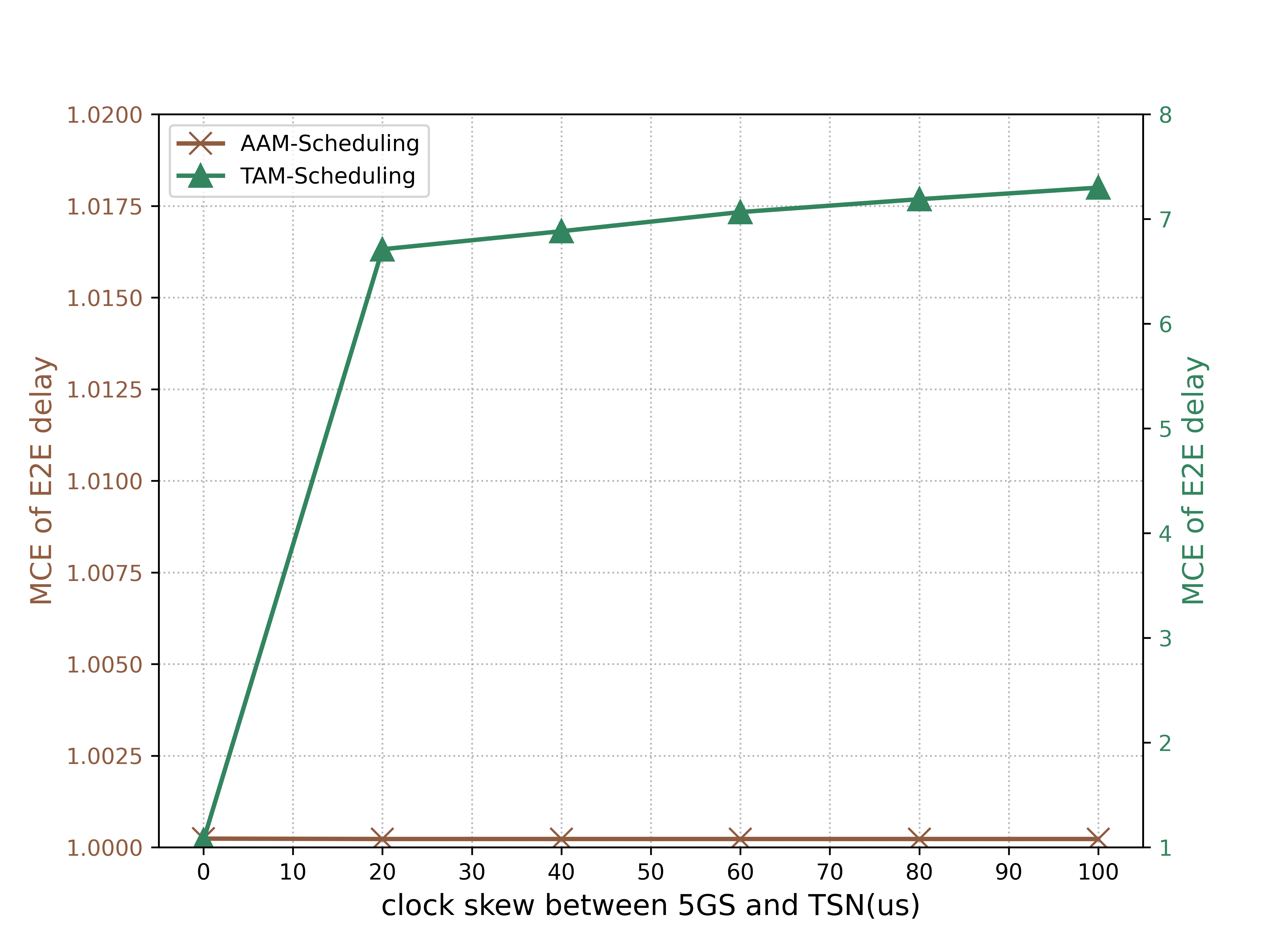}
\caption{The mean coefficient of expansion (MCE) of e2e delay 
varies with clock skew between 5GS and TSN.}
\label{exp-AAM-4}
\end{figure}

\begin{figure}[!t]
\centering
\includegraphics[width=3.5in]{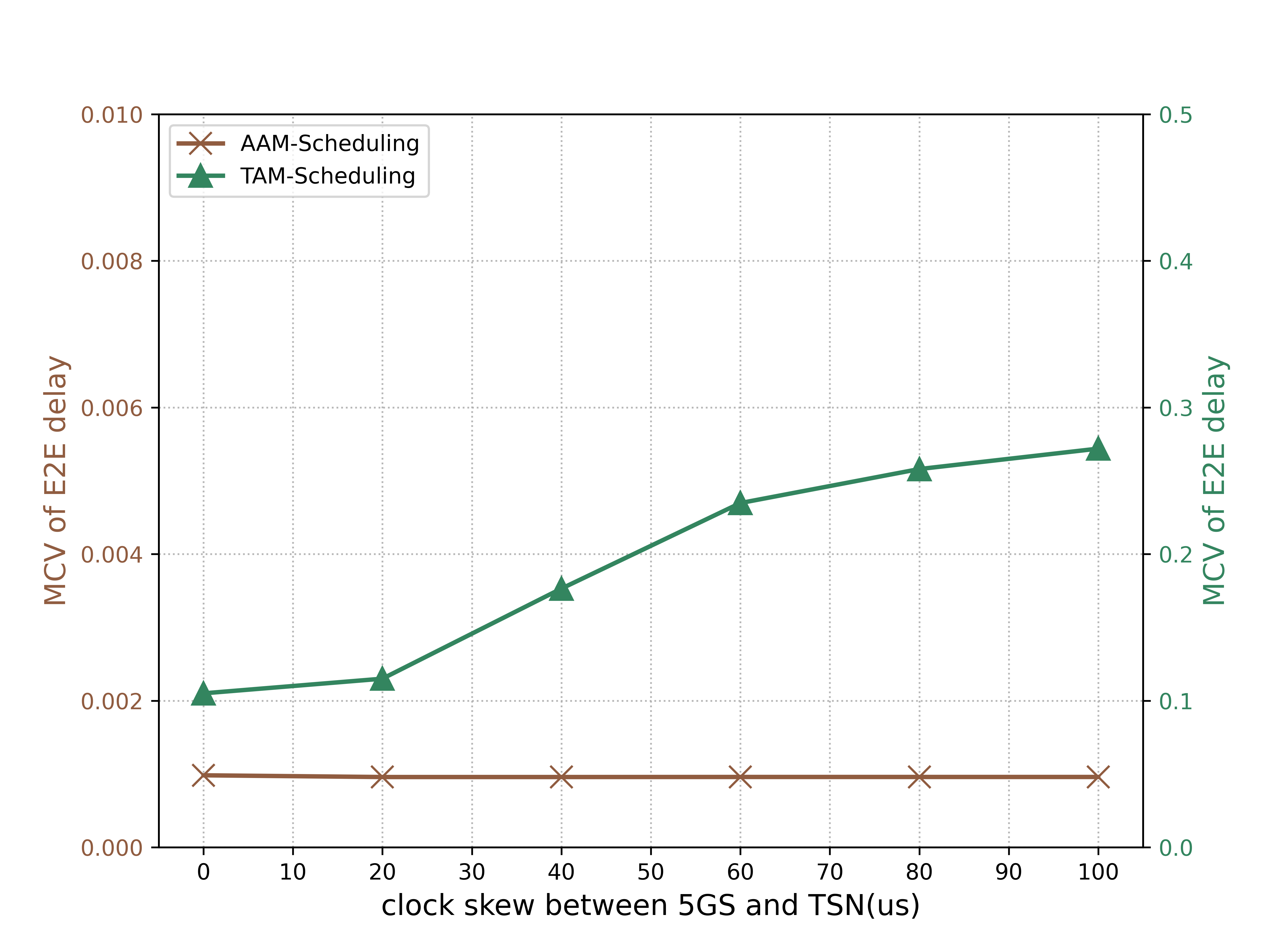}
\caption{The mean coefficient of variance (MCV) of e2e delay 
varies with clock skew between 5GS and TSN.}
\label{exp-AAM-5}
\end{figure}

We keep the 5G transmission jitter fixed and introduce different 
clock skew between the two systems. Specifically, the clock skew is 
changed from $20us$ to $100us$ with step $20us$ and the clock 
deviation follows a uniform distribution. For example, if the clock 
skew equals $20us$, the deviation between the clocks of TSN and 5G 
is sampled from the uniform distribution $U(-10us,10us)$, and 
the positive values indicate that the clock of TSN is earlier than 
5GS, and vice versa. The experiment results are plotted in 
Figs.~\ref{exp-AAM-4} and \ref{exp-AAM-5}. 

As shown in Fig.~\ref{exp-AAM-4}, the MCE of the TAM-Scheduling 
increases as the clock skew increases, consistent with the situations 
shown in Fig.~\ref{fig:AAM-1}. And the MCE of the AAM-Scheduling 
is fixed which is consistent with the situation shown in 
Fig.~\ref{fig:AAM-3}. Fig.~\ref{exp-AAM-5} shows similar results. 
Thus, the AAM can significantly mitigate the delay and jitter 
enlargement caused by the clock skew between the two systems, 
indicating the converged network can still work normally without 
time synchronization.

\subsection{Evaluation of the ATSM}

In the second set of experiments, we evaluate the performance of the 
ATSM in the resource assignment under different traffic loads and 
optimization preferences. All flows have the same parameters and the 
length is $200B$ while the period as well as the delay requirement 
are $1ms$. We then change the number of flows to denote different 
traffic loads and change the $\gamma$ to denote different 
optimization preferences. 

We then calculate TSN resource usage and 5GS resource usage, where 
5GS resource usage is the ratio of the RBs allocated to the TT flows 
to the whole number of RBs in 5GS. In addition to the two types of 
resource usage of the ATSM, TSN resource usage of the STSM is also 
computed to represent the theoretical limit of TSN resource usage 
that can be achieved by the ATSM.

\begin{figure}[!t]
\centering
\includegraphics[width=3.5in]{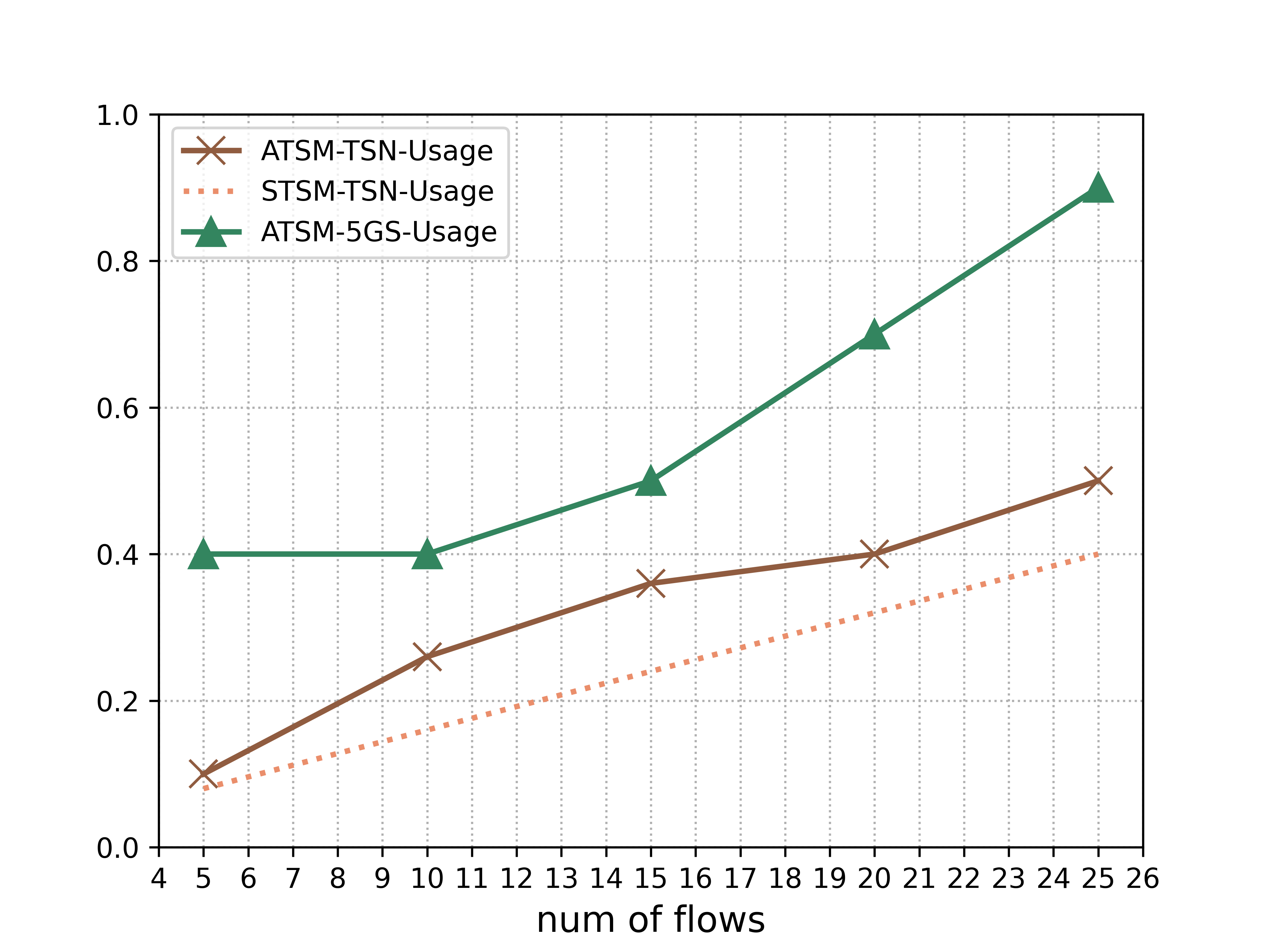}
\caption{TSN and 5GS resource usage scheduled by the ATSM 
with different numbers of flows.}
\label{exp2-1}
\end{figure}

\begin{figure}[!t]
\centering
\includegraphics[width=3.5in]{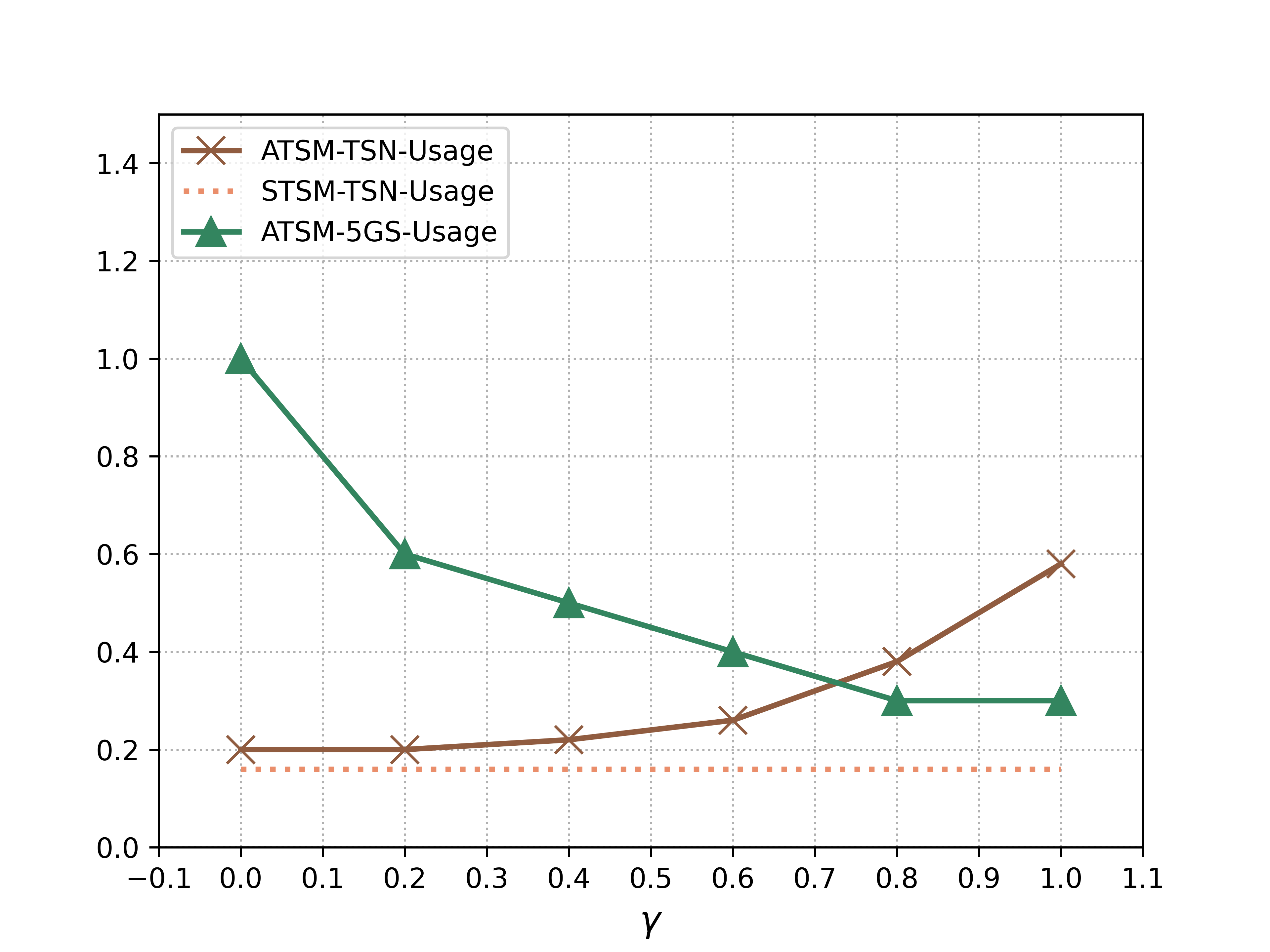}
\caption{TSN and 5GS resource usage scheduled by the ATSM with 
different $\gamma$.}
\label{exp2-2}
\end{figure}

\subsubsection{Performance under different traffic loads}
First, the number of flows is changed from 5 to 25 with step 
size = 5 while other parameters are fixed and $\gamma = 0.5$. 
The relevant results are plotted in Fig.~\ref{exp2-1}.

The resource usage of both 5GS and TSN increases with the number 
of flows. When the traffic load is low (5 flows), TSN resource 
usage is the main resource bottleneck of the whole system. 
In order to reduce TSN resource usage, the ATSM tries to allocate 
larger resource periods to these flows in TSN, thus decreasing
the delay budget in 5GS. To guarantee the delay in 5GS, many RBs are 
allocated by the ATSM, resulting in the low TSN resource usage and 
high 5GS resource usage. As the load increases (10 and 15 flows), 
wireless resource rapidly becomes scarcer. In order to reduce the use 
of RBs in 5GS, ATSM will arrange more flows to be sent in more TTIs 
and fewer RBs, where the delay budgets in 5GS increase. The model 
will allocate smaller resource periods to some flows in order to meet 
their delay requirements. Thus, the slope of TSN resource usage of 
the ATSM is steeper than that of the STSM. When the number reaches 
20 and 25, the RBs used in 5GS cannot be reduced too much in order 
to meet the constraints, so ATSM will allocate larger resource 
periods if possible, which decreases of the steepness of the ATSM
slope. TSN resource usage allocated by the ATSM is shown to be 
always higher than that of the STSM.
The ATSM can thus keep a good balance between TSN and 5GS 
resource usage under different traffic loads.

\subsubsection{Performance under the different optimization preferences}
We now change the $\gamma$ from $0$ to $1$ with step size = 0.2 
and the number of flows is $10$. The relevant results are 
plotted in Fig.~\ref{exp2-2}.

The trend of the two curves is that as $\gamma$ increases, the 5GS 
resources usage decreases while the TSN resource usage increases. 
When $\gamma$ changes from $0$ to $0.2$, the 5GS resource usage 
experiences a sudden decrease since the ATSM start to take care of 
it, and an opposite phenomenon occurs to the TSN resource usage 
when $\gamma$ changes from $0.8$ to $1$. Then, the two resource 
usages gradually increase or decrease but the absolute slope of TSN 
resource usage is less steeper than that of 5GS. This is because 
that the linearization method introduced in Section 
\ref{sec:linearization} provides little room for the ATSM to 
optimize the TSN resource usage, so it is easier to optimize 
that of 5GS.
The ATSM can thus use $\gamma$ to apply different optimization preferences but it's easier to optimize the 5GS resource usage.

\section{Conclusion}
\label{conclusion}
To mitigate the delay and jitter enlargement when 5GS and TSN are 
integrated by TAM, we have presented a new mechanism, AAM, 
for this integration. With this mechanism, the jitter of the e2e 
transmission is isolated only in 5GS and packets from 
5GS to TSN can be delivered at any time without time 
synchronization at the cost of wired bandwidth. 
To coordinate the allocation of resources in the converged network, 
we have then proposed a scheduling model called 
ATSM based on the AAM. Finally, the performance of the AAM and 
the ATSM have been corroborated via simulation using the OMNET++ 
simulator.

As the AAM trades resources for the ability to enable 
asynchronous access for traffic, in future we will consider how 
to dynamically utilize the allocated but un-utilized 
resources effectively.

\ifCLASSOPTIONcaptionsoff
  \newpage
\fi



\bibliographystyle{IEEEtran}
\bibliography{bibtex/bib/IEEEabrv,bibtex/bib/IEEEexample}
%



%

\begin{IEEEbiography}[{\includegraphics[width=1in,height=1.25in,clip,keepaspectratio]{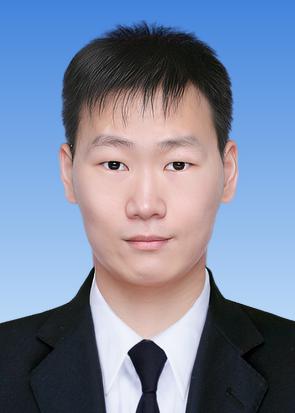}}]
{Jiacheng Li} received the B.S. degree in telecommunication engineering from the
School of Electronic and Information Engineering, Beijing
Jiaotong University(BJTU), Beijing, China, in 2021, where he is currently pursuing his M.S. degree.

His research interests include industrial real-time networking and traffic scheduling.
\end{IEEEbiography}
\vspace{-7 mm}

\begin{IEEEbiography}
[{\includegraphics[width=1in,height=1.25in,clip,keepaspectratio]{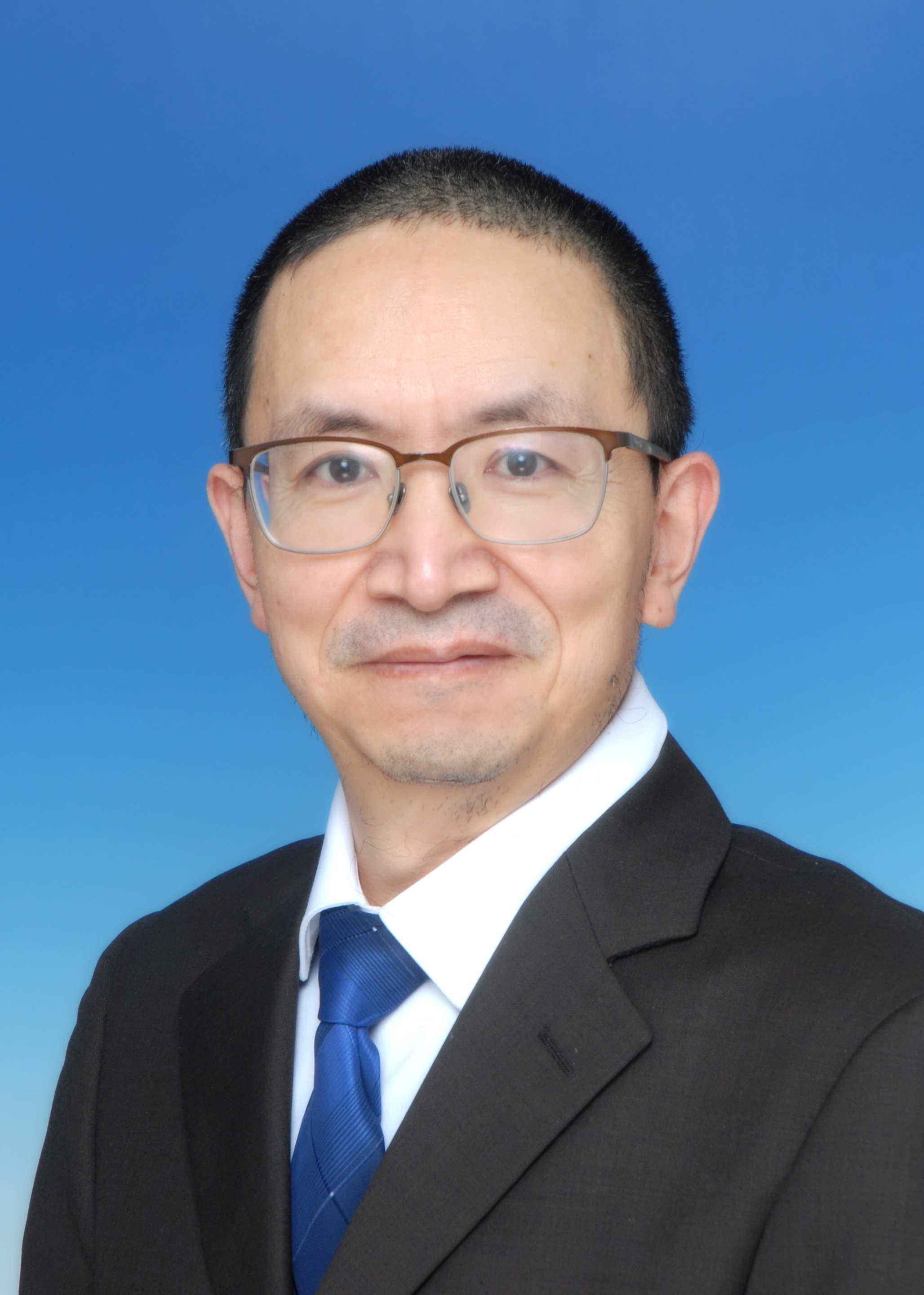}}]{Yongxiang Zhao}
received the Ph.D. degree in electrical engineering from BJTU, Beijing, China, in 2002. 

He is currently an Associate Professor with
the School of Electronic and Information Engineering,
BJTU. His research interests include network protocol and algorithm design, quality of video streaming, and modeling and design of communication networks.
\end{IEEEbiography}
\vspace{-7 mm}


\begin{IEEEbiography}[{\includegraphics[width=1in,height=1.25in,clip,keepaspectratio]{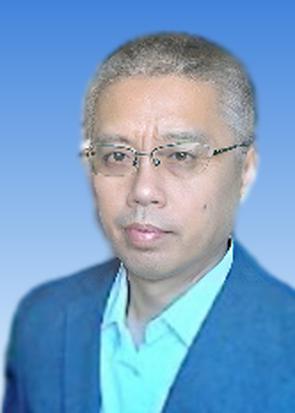}}]{Chunxi Li}
received the B.S. degree in telecommunication engineering, and the M.S. and Ph.D. degrees in communication and information systems from Beijing Jiaotong University (BJTU), Beijing, China, in
1992, 1997, and 2010, respectively.

He is currently an Associate Professor with the
School of Electronic and Information Engineering,
BJTU. His research interests include measurement
and modeling on online mobile streaming systems,
distributed caching network, and industrial traffic scheduling.
\end{IEEEbiography}
\vspace{-7 mm}

\begin{IEEEbiography}[{\includegraphics[width=1in,height=1.25in,clip,keepaspectratio]{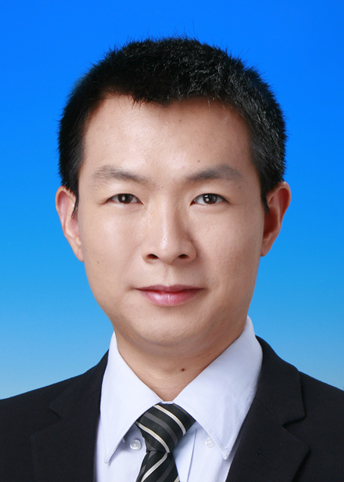}}]{Zonghui Li} received the B.S. degree in computer science from the Beijing Information Science and Technology University in 2010, and the M.S. and Ph.D. degree from the Institute of Microelectronics and the School of Software, Tsinghua University, Beijing, China, in 2014 and 2019, respectively.

He is currently an associate professor in the School of Computer and Information Technology, Beijing Jiaotong University, Beijing, China. His research interests include embedded and high performance computing, real-time embedded systems, especially for industrial control networks and fog computing.
\end{IEEEbiography}
\vspace{-7 mm}

\begin{IEEEbiography}[{\includegraphics[width=1in,height=1.25in,clip,keepaspectratio]{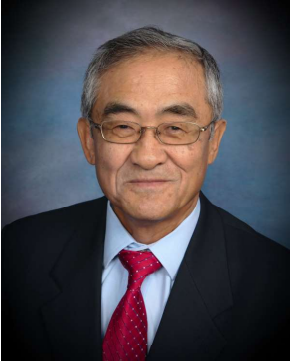}}]{Kang G Shin (Life Fellow, IEEE)} received the BS degree in electronics engineering from Seoul National University, Seoul, Korea, and the MS and PhD degrees in electrical engineering from Cornell University, Ithaca, New York, in 1970, 1976, and 1978, respectively. He is the Kevin \& Nancy O’Connor Professor of Computer Science in the Department of Electrical Engineering and Computer Science, The University of Michigan, Ann Arbor. His current research focuses on QoS-sensitive computing and networking as well as on embedded real-time and cyber-physical systems.

He has supervised the completion of 91 PhDs, and authored/coauthored close to 1,000 technical articles, a textbook and about 60 patents or invention disclosures, and received numerous awards, including the 2022 IEEE Computer Society Technical \& Conference Activities Board Distinguished Leadership Award in real-time systems. He is also an ACM Fellow.
\end{IEEEbiography}
\vspace{-7 mm}

\begin{IEEEbiography}[{\includegraphics[width=1in,height=1.25in,clip,keepaspectratio]{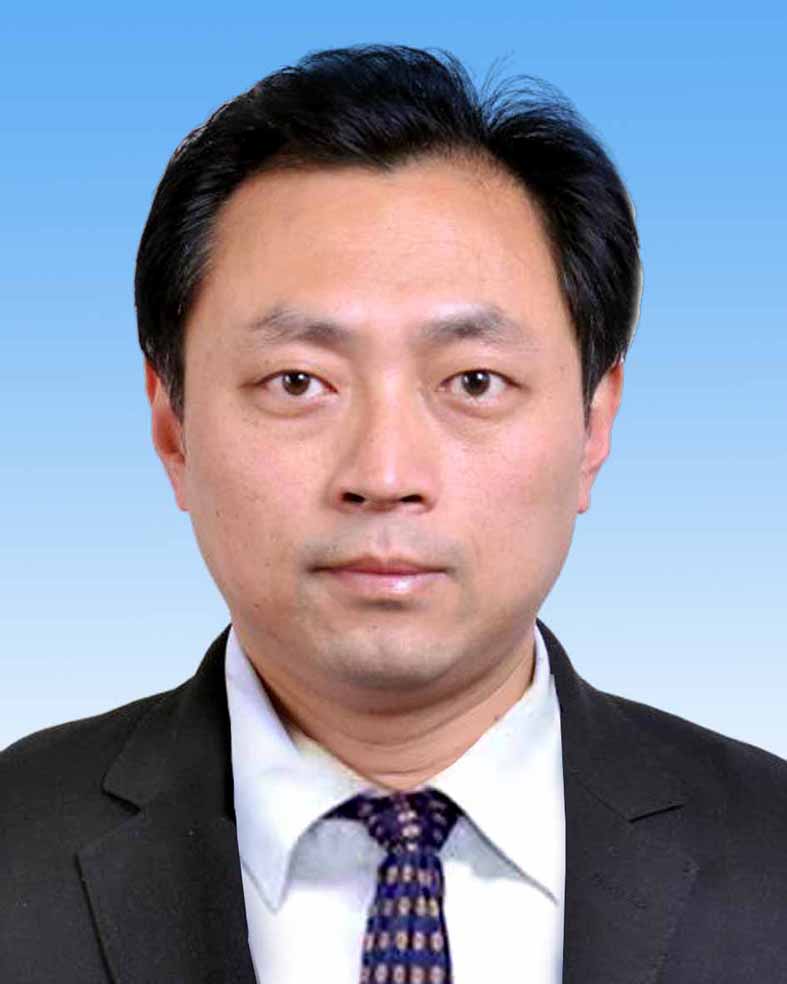}}]{Bo Ai (Fellow, IEEE)} received his Master degree and Ph. D. degree from Xidian University in China. He graduated from Tsinghua University with the honor of Excellent Postdoctoral Research Fellow in 2007. He was a visiting professor at EE Department, Stanford University in 2015. He is now working at Beijing Jiaotong University as a full professor and Ph. D. candidate advisor. He is the Deputy Director of State Key Lab of Rail Traffic Control and Safety, and the Deputy Director of International Joint Research Center.

He has authored/co-authored 8 books and published over 300 academic research papers in his research area. He has hold 26 invention patents. He is a Fellow of the Institution of Engineering and Technology (IET Fellow), IEEE VTS Distinguished Lecturer. He has received many awards including Distinguished Youth Foundation and Excellent Youth Foundation from National Natural Science Foundation of China.
\end{IEEEbiography}




\end{document}